\documentclass[conference,compsoc]{IEEEtran}

\usepackage[dvipsnames]{xcolor}

\usepackage{booktabs}

\usepackage{tabularx}

\usepackage{graphicx}

\usepackage{tcolorbox}
\tcbuselibrary{skins,breakable}

\usepackage{multirow}

\usepackage[hyphens]{url}
\makeatletter\g@addto@macro\UrlSpecials{\do\f{\mbox{f}}}\makeatother

\makeatletter
\let\paper@makecaption\@makecaption
\long\def\@makecaption#1#2{%
  \ifx\@captype\@IEEEtablestring
    \vskip6pt
    \begingroup
      \@parboxrestore\normalfont\footnotesize
      \noindent #1:\enspace #2\par
    \endgroup
  \else
    \paper@makecaption{#1}{#2}%
  \fi}
\makeatother

\definecolor{accent}{HTML}{1F3D63}

\newcommand{\op}[1]{\textsc{#1}}

\usepackage{xspace}
\newcommand{\sysname}{\textsc{Intact}\xspace}
\newcommand{\repname}{\textsc{GenTree}\xspace}
\newcommand{\vanilla}{\textsc{Vanilla}\xspace}

\newtcolorbox{takeaway}{enhanced, breakable, colback=black!4, colframe=black!4,
  boxrule=0pt, arc=1.5pt, boxsep=0pt, left=6pt, right=6pt, top=4pt, bottom=4pt,
  before skip=7pt, after skip=7pt, borderline west={1.2pt}{0pt}{accent}}

\begin{document}

\title{What It Is, Not What It Says:\\ Securing Agentic Browsing Interactions with Web Semantics}

\author{\IEEEauthorblockN{Umar Iqbal}
\IEEEauthorblockA{Washington University in St. Louis}}

\maketitle

\thispagestyle{plain}
\pagestyle{plain}

\begin{abstract}
AI agents increasingly browse the web on users' behalf, yet they consume webpages as screenshots or text, exposing their LLMs to unvetted third-party content.
Browsers enforce semantics under which what an element can do depends on what it is and where it comes from, not on what it says.
Agents flatten these semantics, so injected text can persuade the LLM to exercise the user's authority for content that has none.
As LLMs remain easy to persuade, the agent's interface to the web must itself ensure the integrity of what the agent reads, interprets, and does.
We propose \sysname, which runs on an unmodified browser.
\sysname builds \repname, a bespoke representation that captures provenance and web semantics.
\sysname passes the page via \repname to the LLM one element at a time, so the LLM decides on each element without any other element's text.
Its harness executes each step only within the capabilities of the step's element and asks the user before acting on third-party content.
On 1{,}270 real webpages, \repname attributes 2.3$\times$ as many elements to third parties as the browser, which misattributes 58.7\% of them to the first party.
On 1{,}091 of these webpages with injected attacks, only 12.8\% of attacks get through \sysname's planner, each at most as a consent request, against 73.1\% for a conventional planner.
Isolation disrupts functionality on only 3.7\% of tasks, and a parallel schedule could remove most of its added latency.
\end{abstract}

\section{Introduction}
\label{sec:introduction}

AI agents increasingly act on users' behalf, and the web gives them an API to nearly every online service.
Agentic browsers (e.g., Comet~\cite{comet}, Claude in Chrome~\cite{claudechrome}, and Gemini in Chrome~\cite{geminichrome}) and frameworks (e.g., Browser Use~\cite{browseruse}) read pages, click, fill forms, and navigate across sites, often within the user's authenticated sessions.

This reach carries a fundamental risk, as web content is unvetted.
Pages are assembled at runtime by parties their operators neither control nor always know.
Third parties appear on 92\% of pages, and the median site among the top thousand includes 66 of them~\cite{almanac-thirdparties}.
As third parties load further third parties, about half of websites render content they did not load themselves~\cite{ikram-chain}.
The web is accordingly among the largest attack surfaces in computing.
Cross-site scripting has ranked among MITRE's two most dangerous software weaknesses every year since 2020, and in 2025 SQL injection joined it at the top~\cite{cwe-top25}.

One would thus expect agents to interface with the web through carefully designed mechanisms.
Instead, they simply consume a page as a screenshot or as text derived from its markup~\cite{seeact,browseruse}, exposing the underlying LLM directly to unvetted content.
LLMs are known to follow instructions embedded in such content, i.e., prompt injection~\cite{greshake-indirect}, and to fall for scams~\cite{scamlexity}.
Model-level defenses reduce but do not eliminate either~\cite{struq,secalign,nasr-adaptive}, and OpenAI concedes that prompt injection, ``much like scams and social engineering on the web, is unlikely to ever be fully `solved'\,''~\cite{openai-hardening}.
Such attacks are already manifesting in practice.
Hidden instructions in Reddit posts and comments have led deployed agents to leak users' one-time login codes and credit card numbers~\cite{brave-comet,li-commercial}, crafted links have made them exfiltrate email and calendar data~\cite{cometjacking}, injected page elements and pop-ups divert web agents at high rates~\cite{eia,popups}, and injections aimed at agents appear on live websites~\cite{unit42-ipi,google-ipi}.

The key issue is that web markup was designed for rendering, not for agentic interfacing.
Browsers invest considerable effort in enforcing its semantics.
Each element has well-defined capabilities, and origins and frames bound what code from one party can do to another~\cite{whatwg-html,rfc6454,site-isolation}.
In the browser, what content can do thus depends on what it is and where it comes from, not on what it says.
Screenshots and text serializations flatten these semantics, leaving natural-language text that affords the LLM neither meaning nor control.
A \texttt{<p>} element cannot issue a network request, yet its text can cause one by persuading the agent, which then acts as a confused deputy~\cite{hardy-deputy}, exercising the user's authority for content that has none.

Considering the well-known limitations of LLMs, we argue that security cannot rest on the model's judgment; the interface between the web and the agent must itself ensure integrity~\cite{schneier-trust,schneier-integrity}.
\emph{Input integrity} requires knowing who created or modified each piece of content the agent reads.
\emph{Processing integrity} requires that no element lead the agent beyond what the element itself affords.
\emph{Output integrity} requires that every action, and the data flow it causes, be warranted by the user's request and permitted by the web's security boundaries.
All three must hold with little loss of functionality.

In this paper, we address this problem with \sysname.
Our key idea is to reuse the semantics that browsers already enforce and extend them with the provenance, capturing both in \repname, a per-element representation of the page through which \sysname mediates what the agent reads, interprets, and does.
For input integrity, \sysname attributes every DOM creation and modification to the responsible script~\cite{adgraph,webgraph} and taints what third parties create or modify.
The browser does not make this distinction out of the box, because it attributes everything a third-party script creates in a page to the first-party context.

For processing integrity, \sysname introduces \textit{virtual isolation}, which extends the browser's enforcement of markup semantics to the agent without any change to the browser's architecture.
\sysname passes content to the LLM one element at a time, so an element's text never reaches the decision on another element, and each action the agent derives stays tied to the element it came from, whose properties \sysname can then enforce.
The \texttt{<p>} above thus cannot cause a request, whatever its text says.

For output integrity, \sysname controls each action based on the provenance, capabilities, effects, and security boundaries of the elements involved.
These checks depend only on the page, so \sysname can enforce them reliably.
The user's intent, however, is not encoded in the page, so \sysname defers to the user for decisions that the page cannot settle, e.g., using a third-party booking widget that the task requires. %
\sysname thus need not recognize injected content as malicious or prevent the agent from being persuaded; it limits what a persuaded agent can do.

Our key contributions are as follows:
\begin{enumerate}[\setlength{\IEEElabelindent}{0pt}]
\item We propose and implement \repname, a bespoke representation for AI agents that records the provenance and web semantics of every element.
On tested webpages, \repname attributes 2.3$\times$ as many elements to third parties as the browser's standard attribution, which misattributes 58.7\% of them to the first party.
\item We operationalize \repname in \sysname, a virtual isolation architecture that runs on an unmodified browser.
\sysname passes the page to the LLM one element at a time and executes each step only within that element's capabilities under web semantics.
What an element says thus cannot change what it can do.
\item We evaluate \sysname against \vanilla, a conventional planner, on 1{,}091 real webpages.
Only 12.8\% of attacks get through \sysname's planner, each at most as a request for user consent, against 73.1\% for \vanilla.
\sysname disrupts functionality on 3.7\% of tasks, and a parallel schedule could significantly reduce its time overhead.
\item To foster follow-up research, we release \sysname's source code and our attack benchmark of 1{,}091 real webpages at \url{https://github.com/llm-platform-security/intact}.
\end{enumerate}

Looking ahead, we see \repname as a substrate for existing agent defenses. %
For example, defenses based on execution isolation~\cite{isolategpt}, control- and data-flow enforcement~\cite{camel,fides}, and permission management~\cite{agent-permissions} all presuppose knowing which inputs are untrusted and what each component can do, structure that webpages do not provide explicitly and that \repname derives automatically.

\section{Background and Motivation}
\label{sec:background}

\subsection{Overview of Agentic Browsers}
AI agents that act on the web on a user's behalf are becoming increasingly prevalent. 
These range from native agentic browsers, such as Perplexity's Comet~\cite{comet} and Dia~\cite{dia}, in which the agent is built into the browser, to non-native agents, such as Browser Use~\cite{browseruse} and SeeAct~\cite{seeact}, that drive a commodity browser through external automation, such as Playwright~\cite{playwright} and the Chrome DevTools Protocol~\cite{cdp}.
Conventional browsers, such as Google Chrome~\cite{chrome-agentic} and Microsoft Edge~\cite{edge-copilot}, are also adding agentic features.

At a high level, agents perceive a page through (i) textual representations derived from the DOM~\cite{whatwg-dom} and the accessibility tree~\cite{wai-aria}, (ii) visual representations of the rendered page, or a combination of both. %
With a textual representation, the agent's harness, i.e., the code around the LLM, attaches an identifier to each element, and the LLM names that identifier in its action, e.g., \texttt{click [1582]} in WebArena~\cite{webarena}.
The harness then performs the action on that element, from within the browser in native agentic browsers or through the automation framework in non-native agents.
With a visual representation, the agent instead determines the on-screen coordinates at which to click or type~\cite{webarena}.

\subsection{Security and Privacy Concerns}
\label{sec:background:representation}
These representations serve the agent's functional needs but give little consideration to security and privacy.
They reuse views built for other consumers, namely screenshots for people, the DOM for scripts, and the accessibility tree for assistive technologies.
Once serialized, a third-party advertisement and the site's own checkout button are both just text.
These representations thus undo three of the web's protections: the isolation between origins, the defenses against content injection, and the human in the loop.

First, the web isolates content from different origins.
The same-origin policy keeps one origin's code from reading or acting on another origin's content~\cite{rfc6454}, and site isolation enforces this boundary with separate processes~\cite{site-isolation}.
A representation that flattens the page into one text or one image erases these boundaries, so the LLM receives a third party's content next to the site's own and has no reliable way to tell them apart.
Once the agent follows instructions in that content, the same-origin policy and cross-origin resource sharing become, in Brave's words, ``effectively useless''~\cite{brave-comet}.

Second, the web has developed principled defenses against content injection.
In cross-site scripting, an attacker injects markup that runs with the page's authority, and the web counters it by sanitizing untrusted input.
Agents recast this attack in a form that sanitization cannot catch.
Injected plain text carries no executable markup, so it passes the sanitizer.
Once an agent reads this text, however, its instructions can direct the agent to perform, with the user's authority, the actions that sanitization denies to injected scripts~\cite{greshake-indirect}.

Third, the web's security model counts on a human behind each action.
Browsers gate sensitive operations, such as opening pop-ups, behind user activation~\cite{whatwg-html}, ask the user before a site may use the camera or location, and show the site's origin in the address bar.
These checks presume that the user performs each action and sees each page, which no longer holds when an agent drives the browser.

These weaknesses are serious because LLMs are now capable enough to act on their own, yet remain easy to mislead.
For example, when researchers asked Comet to buy an Apple Watch from a fake Walmart store, in some runs, Comet filled in the saved address and card details without asking for confirmation and completed the purchase within seconds~\cite{scamlexity}.
LLMs also err without an attacker~\cite{webarena}, and model-level defenses make them harder to persuade but not reliable~\cite{struq,secalign,nasr-adaptive}.
The agent's integrity therefore depends on oversight of what goes into the LLM, what the LLM can do with it, and which actions it proposes as a result, i.e., of the LLM's input, processing, and output~\cite{schneier-trust,schneier-integrity}.
This oversight must happen outside the LLM, in code that page content cannot persuade.

\subsection{Key Idea and Challenges}
Our key idea is that the browser already has the information that this oversight needs.
The browser parses every element and enforces its semantics, e.g., a link navigates and a form submits its fields, whereas a \texttt{<p>} element can do neither~\cite{whatwg-html}.
It knows the origin and frame of every document~\cite{rfc6454}, and it runs every script that creates or modifies an element.
We capture this information as the page loads and expose it to the agent's harness in \repname, a bespoke representation for agentic interactions that records, for every element, who created or modified it and what it can do.
We do not rely on websites to author such a representation, because sites adopt even established security mechanisms slowly and incompletely, as with Content Security Policy~\cite{roth-csp}.
Building \repname and using it for oversight raise three challenges, one each for the LLM's input, processing, and output.

\subsubsection{Input Integrity}
The first challenge is to know who created or modified each element, which input integrity requires.
Under our threat model (\S\ref{sec:threatmodel}), injected instructions enter the page through a third party, such as a script, frame, or feed.
A defense can therefore track this untrusted input as taint rather than inspect content for malice, as defenses against cross-site scripting do~\cite{stock-domxss}.
Out of the box, however, the browser attributes what a third-party script writes into the first-party document to the first party, and this standard attribution misses 58.7\% of third-party elements on the pages that we crawled (as we find in \S\ref{sec:eval:attribution}).
Defenses that build on this attribution inherit its gap.
Chrome's agent, for example, hides frames whose origins are unrelated to the task~\cite{chrome-agentic}, but content that a third-party script writes into the first-party document carries the first party's origin and passes this check, so the agent treats it as the site's own.
Recent research defenses do not record authorship either~\cite{ucm, prismata}.
For example, Untrusted Content Masking (UCM) has the site owner, or an LLM, label the regions that hold untrusted content~\cite{ucm}, and Prismata has an LLM infer which kind of party authored each element, e.g., the site's developer, its users, or an external party, from cues in the markup~\cite{prismata}.
Content that a third-party script writes into the first-party document thus counts as the site's own unless a label or a cue gives it away, so these defenses address input integrity only in part.
Further, a site owner cannot label all such content, because webpages are dynamic.
That is, third-party scripts create and modify elements at runtime, and the owner neither controls nor always knows what they write or load.
Moreover, labels that an LLM infers rest on the LLM's judgment, which brings back the very problem that input integrity should solve.
We instead build on prior work on browser instrumentation that records how scripts construct a page~\cite{adgraph,webgraph,pagegraph}, attributing every DOM creation and modification to the responsible script and tainting what third parties create or modify, without relying on any LLM (\S\ref{sec:design:instrumentation}).

\subsubsection{Processing Integrity}
The second challenge is to keep the content that the LLM reads from steering how the LLM processes the rest of the page.
When the LLM reads the whole page at once, one element's text can direct an action on any other element, and the agent becomes a confused deputy~\cite{hardy-deputy}.
The attacker needs such an action, because its goal must exceed what its own element can do, or it would not need the agent.
Prior defenses, such as UCM~\cite{ucm} and Prismata~\cite{prismata}, still pass the page to one LLM that plans over it as a whole, after they restrict the content that they consider untrusted, e.g., user reviews and advertisements.
Whatever the labels miss thus still reaches an LLM that sees the whole page, and a single mislabeled element raised UCM's attack success rate from 0\% to 6\%, close to the 8\% of a prompt-based defense~\cite{ucm}.
Further, these defenses restrict the capabilities of \textit{untrusted} content, whereas we find that third-party content provides useful functionality, such as links, forms, and text fields (Table~\ref{tab:rq1-capabilities}).

We instead plan one element at a time, third-party elements included, and the call that decides an element sees no other element's text (\S\ref{sec:design:planning}).
Each call can thus propose steps only on its own element, because our code, not the LLM, ties each proposed step to that element.
Planning one element at a time, however, raises a challenge of its own, because a step often depends on earlier steps, e.g., after a query is typed, the next step is the search button.
To address this, we pass a log of the operations decided so far, without the text of any other element.
Our intuition is that a capable LLM needs to know only what the plan has done so far, not what other elements say.
Still, this limited context can disrupt some tasks, so we measure how much functionality isolation costs (\S\ref{sec:eval:functionality}).

\subsubsection{Output Integrity}
The third challenge is to decide what the agent may do on each element.
The harness executes a step only if, under web semantics, the element affords the step's operation.
For example, the harness rejects a click on a heading that sits outside any link or button, whatever the heading says.
Deriving these capabilities is non-trivial, because neither the tag nor the accessibility tree always shows what an element does.
An ARIA role can turn a \texttt{<div>} into a button, script rather than the tag defines a custom element's behavior, and a hidden input is absent from the accessibility tree yet is submitted with its form.
To address this, we build a taxonomy of capabilities that follows the HTML specification~\cite{whatwg-html} and ARIA~\cite{wai-aria}, i.e., the interactions that each element affords and the effects that they can have, and we label every element of \repname with it, accounting for the element's ancestors, ARIA attributes, and computed style (\S\ref{sec:design:gentree}).
Further, web semantics cannot settle every step, e.g., whether to use a third-party booking widget that the task requires, so some steps need the user's consent.
Asking about every step would defeat the purpose of an agent, but provenance and web semantics let the harness ask only about the riskiest steps, e.g., steps on third-party elements or steps with external effects. %
Together, these answers improve the security of the agent by design rather than by the LLM's judgment.

\subsection{Complexity of Real Webpages}
We design \sysname and \repname to work on real webpages in practice, which are complex and which benchmarks struggle to replicate. %
For example, many parties assemble a real page at runtime, they load further parties, embed services such as payment forms in cross-origin frames, and modify one another's elements. %
Self-hosted pages lack the live third parties that make attribution hard and that tasks rely on, and a few sites cannot show whether a defense holds across the variety of the web.
We therefore evaluate \sysname on real webpages captured with their live third-party content, 1{,}270 for attribution and 1{,}091 for attacks and functionality (\S\ref{sec:evaluation}).
We also release the 1{,}091 webpages with our attacks as a benchmark.

\section{Threat Model}
\label{sec:threatmodel}

\vspace{2mm}
\noindent
\textbf{System Model.} We consider an agent that autonomously interfaces with the web. 
The agent consists of an LLM and a harness, i.e., the code around the LLM.
The harness gives the LLM a view of the page, e.g., its markup, and executes the LLM's decisions, such as clicking elements, filling in forms, and navigating.
Webpages combine first-party content with third-party scripts, advertisements, and embedded services, which may appear directly in the first-party document or in cross-origin frames.
Third-party content may originate from arbitrary sources and is not known to the user or the first-party operator in advance.

\vspace{2mm}
\noindent
\textbf{Attacker Capabilities and Goals.} We assume a third-party attacker who controls \emph{some} content on a page that the agent visits, but has no special access to the user, the agent, or the first-party origin.
The attacker may control only content, e.g., a promotional note that an honest third party's script relays into the first-party document.
It may run code in its own cross-origin frame, e.g., an embedded payment form.
The attacker's content may be hidden from view and still reach the agent.

The attacker may target an element of the first party or of another third party (\emph{outbound} attacks) or present its own elements as bait (\emph{inbound} attacks).
We consider attacks that require the agent, i.e., attacks that the attacker cannot carry out with its own capabilities.
A \texttt{<p>} element cannot carry out such an attack; a script can, unless it runs inside a cross-origin frame.
We impose this condition because an attacker who can achieve its goal through its own scripts or content does not need the agent.

\vspace{2mm}
\noindent
\textbf{Trust Relationships.}
We assume that the user, browser, and the agent harness are trustworthy and uncompromised.
The LLM has no intent of its own to harm the user, but we do not trust its judgment.
We assume that the attacker can persuade it to follow any instruction in the content that it reads, since model-level defenses reduce but do not eliminate prompt injection~\cite{struq,secalign,nasr-adaptive}.
The LLM may also err without an attacker~\cite{webarena,iqbalAIES24}.
We assume an honest first party and trust what its own responses and scripts deliver.
We treat as untrusted all content in third-party frames, every element that a third party's script creates or modifies, and every write whose script cannot be identified.
Our instrumentation runs in the page's JavaScript environment, so we also assume that page scripts do not evade or tamper with it; moving the instrumentation into the browser would remove this assumption (\S\ref{sec:discussion}).

\vspace{2mm}
\noindent
\textbf{Scope.}
Our goal is to prevent the attacker's content from leading the agent to act beyond the user's request, whether the content sits in the first-party document or in a cross-origin frame.
Concretely, the attacker's text may reach the decisions on its own elements but no decision on another element.
Every step must stay within its element's capabilities, and no step on a third-party element may execute without the user's consent.
We consider the webpage on which the agent plans, not attacks that span several pages.

\noindent
\textit{Out of scope.} We do not seek to detect or eliminate prompt injection; we aim to constrain its effects when the agent encounters untrusted content.
We also exclude attackers who control the first-party origin itself.
We do not address availability: an attacker can make the agent fail a task, e.g., with a decoy that draws the planner away from the part of the page that the task needs (\S\ref{sec:discussion}).

\section{\sysname: System Architecture}
\label{sec:design}

We propose \sysname, an execution architecture for secure agentic web interfacing.
\sysname's main thesis is that the agent's integrity depends on oversight of the LLM's input, processing, and output, as the LLM itself remains unreliable.
\sysname bases this oversight on web semantics, i.e., on what each element is and where it comes from, not on what the element says.
To oversee the input, \sysname builds \repname, a bespoke representation of the page that records, for each element, which party created or modified it and what it can do (\S\ref{sec:design:gentree}).
To oversee the processing, \sysname plans under virtual isolation (\S\ref{sec:design:planning}).
It shows the LLM one element of \repname at a time instead of the whole page, so the LLM decides on each element without any other element's text.
The resulting plan carries little information beyond what the user asked for, yet it is directly actionable.
To oversee the output, the harness (not the LLM) turns the plan into browser actions.
It permits each step only within the capabilities of the element that the step names, and it asks the user before it acts on any third-party element.

\begin{figure*}[t]
\centering
\includegraphics{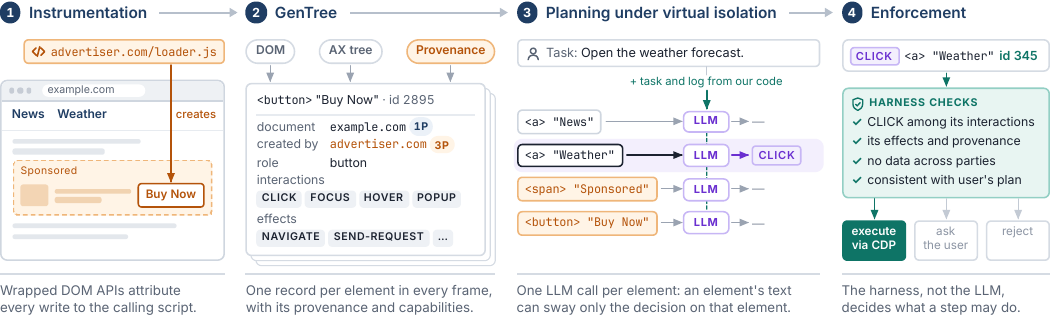}
\caption{Overview of \sysname. (1) As the browser loads the page, our instrumentation attributes every creation and modification of an element to the script responsible for it and taints what third parties create or modify (\S\ref{sec:design:instrumentation}). (2) \sysname then combines the DOM of every frame, the accessibility tree, and these attributions into \repname, labeling each element with its capabilities (\S\ref{sec:design:gentree}). (3) To plan a task, \sysname shows the LLM one element at a time and records the operation, if any, that the task needs on that element (\S\ref{sec:design:planning}). (4) The harness executes the resulting plan, permitting each step only within the capabilities of the element it names. Amber marks third-party content, teal marks what our code adds, and violet marks the LLM.}
\label{fig:overview}
\end{figure*}

Figure~\ref{fig:overview} describes the execution flow in \sysname.
We implement \sysname in Python on stock Chromium, which we drive through Playwright~\cite{playwright} and the Chrome DevTools Protocol (CDP)~\cite{cdp}.
We release our source code at https://github.com/llm-platform-security/intact.

\subsection{\repname Construction}
\label{sec:design:gentree}

To control what the LLM reads and what its decisions can cause, \sysname needs a representation of the page in which this control can be expressed.
Agents currently reuse representations built for other consumers, namely screenshots, the DOM, and the accessibility tree, which lend no such control (\S\ref{sec:background:representation}).
We create \repname, which extends these generic representations with the semantics that \sysname needs, so that each element carries what the harness requires to decide what that element's content may cause.

\begin{figure}[t]
\centering
\includegraphics{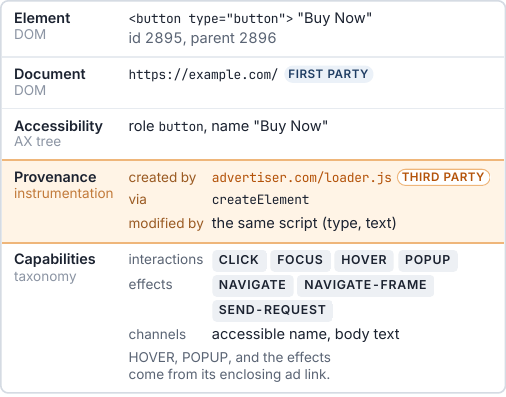}
\caption{An abridged \repname record from our crawl, with its domains anonymized. An advertiser's script created this button in the site's first-party document.}
\label{fig:gentree-record}
\end{figure}

\repname combines the DOM of every frame, the accessibility tree, and provenance into one record per element, which makes it a much richer representation than any of the three alone (Figure~\ref{fig:gentree-record}).
Each record holds the element's tag, attributes, direct text, and parent; the document it belongs to and whether that document is first- or third-party; its accessibility role and name, when the element is in the accessibility tree; its provenance, i.e., the script that created it, if any, and the parties that modified its attributes, text, or event listeners; and its capabilities.\footnote{We identify a party by its registrable domain (eTLD+1, computed with the Public Suffix List~\cite{psl}).}
\repname also keeps the elements that the accessibility tree omits, such as scripts, hidden inputs, and hidden subtrees, because their content can still reach an agent.
As each record already carries the element's provenance and capabilities, \repname also provides extensibility for fine-grained access control that may encode user preferences over them.

\subsubsection{Web Semantics Representation}
Provenance tells \sysname who created or modified an element, but not what that element can do.
\repname thus also records each element's capabilities under the web's semantics, so that the harness can oversee the output.
To that end, we create a taxonomy of capabilities along two axes, following the semantics that the HTML specification~\cite{whatwg-html} and ARIA define for each element: (i) the \emph{interactions} that the element affords an agent, e.g., \op{click}, \op{type}, and \op{upload} (ten in total), and (ii) the \emph{effects} that these interactions can have in the browser, e.g., \op{navigate}, \op{send-request}, and \op{dom-mutate} (thirteen in total).
Effects are limited to observable browser consequences; whether a request amounts to, say, a purchase is a judgment that we leave to a separate layer.
We statically assign these labels from each element's tag and attributes with 76 rules, grouped by element family such as links, forms, media, embedded content, and dialogs, and 14 composition rules then adjust them for the element's ancestors, ARIA attributes, and computed style.
For example, text inside a link inherits the link's \op{click}, and a disabled button loses all its interactions.
\repname also records the channels through which an element's content reaches the agent, e.g., its text, accessible name, or placeholder.

The taxonomy further records whether each effect can be undone.
A \emph{local change}, such as \op{dom-mutate}, can be reverted; an \emph{external effect}, such as \op{send-request}, cannot be recalled once it occurs; and an \emph{undetermined} effect, such as \op{submit} or \op{navigate}, has consequences that the label alone does not settle (Appendix~\ref{sec:effect-reversibility} lists the category of each effect).
Typing into a field of the first-party document is a local change, or undetermined when the field sits in a form, since typing can submit the form.
Typing into a field of a third party's frame is an external effect, because that third party can read the value.
Custom elements, whose behavior is defined by script rather than by the specification, cannot be labeled statically.
\repname exposes their text but grants them no interactions unless an ARIA role declares one, e.g., \texttt{role="button"} confers \op{click}; the harness therefore does not act on them otherwise, which favors safety over functionality.

\subsubsection{Browser Instrumentation}
\label{sec:design:instrumentation}

\repname's provenance requires knowing which script, if any, created each element and which scripts modified its attributes, text, or event listeners, but the DOM records neither.
The browser attributes everything that a script writes into a document to that document's origin, so content that a third-party script injects into the first-party page is indistinguishable from the site's own, as with the button in Figure~\ref{fig:gentree-record}.
We therefore instrument the browser to attribute every DOM creation and modification to the responsible script, which lets \repname separate what the site wrote from what a third party injected into it.
Before the page loads, we use CDP to register a script that runs in every frame before the page's own scripts, wraps the DOM interfaces through which scripts create, insert, and modify elements, and attributes each call to the running script, which it reads from the JavaScript call stack.

\vspace{2mm}
\noindent
\textbf{Script Identity.} We identify a script by its party, its URL, and the hash of its content, which CDP reports for every script that the JavaScript engine compiles, so that different scripts served from the same URL remain distinct.
Code that has no URL of its own, i.e., code passed to \texttt{eval} or \texttt{Function} and inline scripts that another script inserts, would hide its origin from the call stack, so we give it a synthetic source URL that maps back to the script that created it; the functions it defines, including callbacks that run later, such as timers, thus stay attributed through any chain of such indirections.
We strip the source URLs that page code declares so that one script cannot pose as another.
A write whose script cannot be identified is marked as unattributed and treated as untrusted rather than as first-party.

\vspace{2mm}
\noindent
\textbf{Coverage.} For each element, the wrappers record the script that created it, the script that inserted it, and the parties that set its attributes or text or attached event listeners to it.
Interfaces that build elements from HTML strings, such as \texttt{innerHTML} and \texttt{document.write}, do not return the elements that they create, so we collect these elements with a mutation observer that we read synchronously around the call.
CDP's network events attribute each third-party request to the script that initiated it.
Cross-origin frames, which Chromium's site isolation runs in separate processes~\cite{site-isolation}, get their own CDP session and the same instrumentation, and are merged into the page's \repname.
The instrumentation does not yet cover cookies, \texttt{postMessage} exchanges, content that browser extensions inject, or shadow roots, and it can be extended to them.
Appendix~\ref{sec:instrumentation-details} gives further implementation details of the instrumentation.

\subsection{Planning under Virtual Isolation}
\label{sec:design:planning}

\repname tells \sysname what each element is, but deciding what a task requires still takes the LLM's judgment.
\sysname therefore needs a planning paradigm that keeps control and integrity with the harness, that the LLM cannot simply evade, and that still lets the agent carry out the user's tasks.
When the LLM reads the whole page at once, as agents do today, one element's text can redirect how the LLM treats every other element, and an action that it proposes cannot be traced to the content that prompted it.
We instead iterate over \repname and show the LLM one element at a time, which, for all intents and purposes, runs each element in isolation: its text can inform the decision about that element but no decision about another, which we refer to as virtual isolation.

\begin{figure}[t]
\centering
\includegraphics{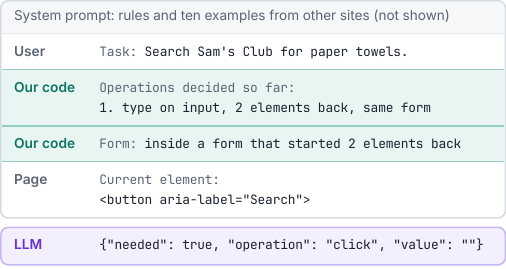}
\caption{One element's call to the LLM, for a task on samsclub.com. Our code, not the LLM, writes the log and the form line.}
\label{fig:planner-call}
\end{figure}

Planning one element at a time raises the question of what context the LLM needs.
Earlier elements are useful context, e.g., after a query is typed, the next step is the search button, but passing their text would reintroduce the cross-element influence that isolation removes.
We therefore pass the minimal context necessary (Figure~\ref{fig:planner-call}):
(i) the user's request;
(ii) the current element's own markup, i.e., its tag, a fixed set of visible attributes (e.g., \texttt{href}, \texttt{aria-label}, \texttt{placeholder}, and \texttt{type}), and its direct text;
(iii) a log of the operations decided so far, each with the tag name of the element it was decided on, how many elements back that element is, and whether it shares the current element's form; and
(iv) a line that says whether the current element is inside a form and how many elements back that form began.
Our code derives the log and the form line from the page's structure, and neither carries text, attribute values, or typed values from other elements; the log reports a custom element's tag, whose name the page chooses, only as \emph{custom}.
The LLM never sees provenance, parties, capability labels, or element identifiers, which are inputs to the harness rather than to the LLM.

The log and the form line tell the LLM where it is in a task without the text of any other element.
In the search example above, the log shows a \op{type} one element back in the same form, so the current element is likely the search button.
The request supplies the query to type.

We keep the operations close to what a browser actually does on a page: it clicks, types, and selects.
For each element, the LLM answers whether the task needs the element and, if so, with which operation, \op{click}, \op{type}, or \op{select}, and the value to type or choose; it answers \op{none} when the task needs only the element's content, e.g., a price that it must compare.
The LLM makes this decision as a classifier guided by few-shot examples in the same format, which we take from pages outside our evaluation (Lowe's, iHerb, Home Depot, and PriceRunner).
The prompt's rules and examples teach patterns that a single element cannot reveal, such as choosing only the element that performs an action rather than the menus and icons around it, and filling only the fields that the task asks for, since a script may have added others to the form.
The LLM never sees or returns element identifiers.
Our code links each decision to the element on which it was made, so the LLM cannot direct an action to any element other than the one it was shown.

Showing the LLM every element of a page would take thousands of calls per task; IMDb's homepage alone has 3{,}355 elements with visible markup.
We therefore split the page's markup, in page order, into chunks of about 1{,}500 characters, and ask the LLM, once per chunk and in parallel, whether the chunk contains anything that the task must act on or read.
If several chunks qualify, one more call shows the LLM those chunks as numbered parts, and it picks the one where the task can best be done.
The LLM then walks that chunk one element at a time, in page order.
We cut chunks by size alone rather than at form boundaries, because a script can inject fields into a form, so a form is not a trustworthy unit.
A task that needs a form, however, needs all of it: when the chosen chunk ends inside a form, the walk continues element by element until the form ends, which is the only way in which the walk crosses a chunk boundary.

Virtual isolation covers the calls that decide on elements, but two things that shape the walk lie outside it: (i) which chunk the walk covers and (ii) the log of earlier decisions.
The relevance and choice calls read whole chunks, so text that an attacker injects can steer which chunk the planner walks, although these calls propose no step.
The log reports the operations decided on earlier elements, including an attacker's element, so the decision on that element can affect how later calls read their own elements, although the log carries none of its text.
Neither input lets an element's text reach the decision on another element or name an element to act on.
Instead, both let an attacker derail the plan, e.g., with a decoy that draws the walk away from the part of the page that the task needs, which affects whether the task completes rather than what the agent may do; Section~\ref{sec:discussion} discusses these attacks on availability.

\begin{figure}[!t]
\centering
\includegraphics{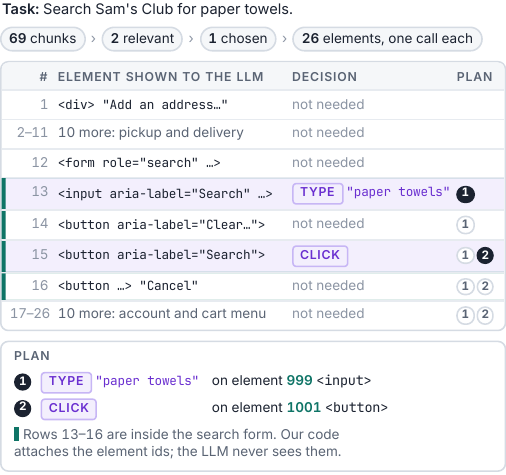}
\caption{Planning the task of Figure~\ref{fig:planner-call} on samsclub.com, one element per call. Our code attaches the element identifiers to the plan.}
\label{fig:plan-construction}
\end{figure}

The output is a plan, which is the ordered steps that the LLM marked as needed, each with the element's identifier, the operation, and the value (Figure~\ref{fig:plan-construction}).
Our code also records, for each \op{click}, the link or button that the click activates, because clicking text inside a link activates the link.
A plan thus carries little information beyond what the user asked for, yet the harness can act on it directly.

The harness, not the LLM, turns the plan into actions.
It executes each step through CDP on the element that the step names, and only if the step's operation is among that element's interactions in \repname: a \texttt{<p>} element has no \op{click} unless it sits inside a link or button, whatever its text says.
It also asks the user before it acts on any third-party element.
As the operation, its effects, and the element's provenance are applied outside the LLM and derive from web semantics, every step comes from the decision on its own element and stays within that element's capabilities. %
\sysname's guarantees thus rest on its design rather than on the LLM's judgment.

\subsubsection{Functionality Extraction}
\label{sec:design:functionalities}

In deployment, the user supplies the request.
To plan and evaluate realistic tasks at scale, we need the requests that real users would make of an agent on each page.
We extract them with an LLM, where one call per page shows the page as markup only, in the same view that the planner uses, and asks for five realistic functionalities, each phrased as the user would ask for it, with concrete values where needed.
We ask for a mix of interactions, namely a single action (one click or choice), several actions (e.g., filling in a form), and reading the page to decide what to do (e.g., finding the cheapest option and opening it), and require every request to end in acting on the page rather than in a summary or explanation.
Each functionality records a short name, the request, and its interaction type.

\section{Evaluation}
\label{sec:evaluation}

We evaluate \sysname along three axes:
(i) whether the standard browser attribution of page content is as limited as we argue, and whether \repname corrects it; %
(ii) whether \sysname stops injected content from acting beyond the user's request and the element's capabilities, even when the agent is fully persuaded, and where it falls short; %
and (iii) how well \sysname's planning carries out functionalities while exposing limited context to the LLM, how much it reduces the decisions left to the user, and what latency and cost its security measures add.

\subsection{Data Collection}
\label{sec:eval:setup}
We evaluate on real webpages rather than on synthetic benchmarks to capture the nuances of the live web, where several third parties may load content, redirect requests, and interact in complex ways with each other and with the first party.
To obtain pages that are both popular and diverse, we sample websites from the Tranco list~\cite{tranco} by category, taking the categories from Curlie~\cite{curlie} and relying on an LLM to resolve multi-category conflicts. %
Specifically, we label each of the top-50K Tranco websites in fifteen topical categories and sample 100 websites per category uniformly at random.
Two categories fall short of 100 websites, Adult (65) and Kids \& Teens (21), because few of the websites listed under them are primarily about these topics, which leaves 1{,}386 websites.

We crawl the landing page of each website in September 2026 with a crawler built on Playwright~\cite{playwright}, which drives Chromium with our instrumentation (\S\ref{sec:design:instrumentation}) and loads each page with its live content.
The crawler lets each page settle for eight seconds before capturing it, retries once with a 20-second wait when the load fails or a bot challenge appears, and skips a page that stalls for more than five minutes.
Of the 1{,}386 pages, 65 failed to load, 45 served a bot block or challenge, and six stalled, which leaves 1{,}270 pages.
On these pages, the crawler also captured the contents of 15{,}854 of the 15{,}968 cross-origin frames. %

\subsection{Extended Attribution via Instrumentation}
\label{sec:eval:attribution}
Input integrity rests on attribution, so we first measure how far \repname extends the browser's standard attribution.
The browser tells which origin content belongs to, whereas the agent needs to know which party wrote it, and the two diverge whenever a third-party script writes into the first-party document.
Such writes break origin-based trust, since an agent that trusts first-party content then also trusts whatever third parties inject into it.
Without instrumentation, an element can be attributed to a third party only if it lies in a third-party frame or loads a resource, such as a script, image, or frame, from a third party.
With instrumentation, \repname also attributes to third parties the elements that their scripts create and the first-party elements whose behavior or content a third party changes, through attributes, text, or event listeners.

\begin{table*}[t]\centering\small
\renewcommand{\arraystretch}{1.15}
\setlength{\tabcolsep}{4pt}
\begin{tabular*}{\textwidth}{@{\extracolsep{\fill}}lrrrrrrrrr@{}}
\toprule
& & \multicolumn{2}{c}{\textbf{\shortstack{\vanilla\\Attribution}}} & \multicolumn{6}{c}{\textbf{\shortstack{\repname\\Instrumentation}}} \\
\cmidrule(lr){3-4}\cmidrule(lr){5-10}
\textbf{Category} & \textbf{Pages} & \textbf{Total 3P} & \textbf{Iframe} & \textbf{Total 3P} & \textbf{Iframe} & \textbf{\shortstack{3P\\Created}} & \textbf{\shortstack{1P\\Modified}} & \textbf{\shortstack{3P\\Modified}} & \textbf{Additions} \\
\midrule
Adult & 62 & 5.6\% & 25.1\% & 16.5\% & 18.0\% & 25.1\% & 10.9\% & 0.1\% & 35.9\% \\
Arts & 93 & 12.3\% & 44.6\% & 25.6\% & 33.3\% & 34.2\% & 6.7\% & 1.1\% & 40.9\% \\
Business & 98 & 7.6\% & 24.5\% & 22.4\% & 14.4\% & 44.3\% & 10.9\% & 1.6\% & 55.2\% \\
Computers & 88 & 11.6\% & 36.8\% & 22.4\% & 25.8\% & 27.1\% & 9.1\% & 1.8\% & 36.2\% \\
Games & 90 & 11.7\% & 36.8\% & 27.1\% & 28.0\% & 38.1\% & 8.8\% & 1.0\% & 46.9\% \\
Health & 90 & 8.8\% & 32.3\% & 19.7\% & 23.1\% & 38.0\% & 14.4\% & 2.1\% & 52.4\% \\
Home & 92 & 13.7\% & 50.6\% & 24.3\% & 37.7\% & 32.8\% & 8.0\% & 1.2\% & 40.8\% \\
Kids \& Teens & 19 & 15.9\% & 62.7\% & 21.3\% & 53.8\% & 21.2\% & 2.9\% & 0.7\% & 24.1\% \\
News & 94 & 18.3\% & 58.4\% & 34.3\% & 42.8\% & 36.5\% & 11.6\% & 1.8\% & 48.1\% \\
Recreation & 91 & 8.3\% & 34.0\% & 19.3\% & 22.3\% & 40.9\% & 8.9\% & 1.5\% & 49.8\% \\
Reference & 86 & 10.1\% & 36.8\% & 24.9\% & 25.3\% & 30.0\% & 18.0\% & 1.1\% & 48.0\% \\
Science & 86 & 6.2\% & 28.5\% & 14.9\% & 19.8\% & 35.4\% & 11.3\% & 1.2\% & 46.7\% \\
Shopping & 95 & 8.3\% & 33.2\% & 22.3\% & 19.3\% & 41.1\% & 8.9\% & 1.8\% & 50.0\% \\
Society & 90 & 4.9\% & 23.2\% & 12.9\% & 17.3\% & 33.5\% & 11.1\% & 0.9\% & 44.6\% \\
Sports & 96 & 10.6\% & 46.9\% & 28.4\% & 31.8\% & 48.1\% & 7.9\% & 1.9\% & 56.0\% \\
\midrule
\textbf{All} & \textbf{1{,}270} & \textbf{10.0\%} & \textbf{37.5\%} & \textbf{22.7\%} & \textbf{26.4\%} & \textbf{36.4\%} & \textbf{10.3\%} & \textbf{1.4\%} & \textbf{46.7\%} \\
\bottomrule
\end{tabular*}
\caption[Third-party attribution without and with instrumentation]{Third-party (3P) elements without and with instrumentation, as per-page means. Total 3P columns represent the percentage of a page's elements that are third-party, and Iframe columns represent the percentage of these elements inside iframes. The other columns represent percentages of the third-party elements with instrumentation. Additions column represents the elements attributed only with instrumentation, which third-party scripts created (3P Created) or which are first-party (1P) elements whose attributes, text, or event listeners third-party scripts changed (1P Modified). 3P Modified column represents third-party elements whose behavior or content a different third party changed.}
\label{tab:rq1-attribution}
\end{table*}

\subsubsection{Results}
Table~\ref{tab:rq1-attribution} presents the third-party content that our instrumentation captures.
The browser's standard attribution assigns only 10\% of a page's elements to third parties on average, whereas \repname assigns 22.7\%, 2.3 times as many.
On the median page with third-party content, 46.5\% of the third-party elements are visible only through instrumentation; across all pages, the browser's attribution misses 58.7\% of third-party elements, and 1{,}095 of the 1{,}270 pages (86\%) contain at least one missed element.
Most of the difference comes from elements that third-party scripts create inside first-party documents, which make up 36.4\% of a page's third-party elements on average.

Third parties also change content that they did not create.
On 836 pages (65.8\%), third-party scripts modify first-party elements, which make up 10.3\% of a page's third-party elements on average.
On 567 pages (44.6\%), a third party changes elements that belong to another third party, which make up 1.4\% of a page's third-party elements on average.

\begin{table}[t]\centering\small
\renewcommand{\arraystretch}{1.15}
\setlength{\tabcolsep}{4pt}
\begin{tabular*}{\columnwidth}{@{\extracolsep{\fill}}lrrr@{}}
\toprule
& & \multicolumn{2}{c}{\textbf{Missed}} \\
\cmidrule(lr){3-4}
\textbf{Capability} & \textbf{\shortstack{3P\\Elements}} & \textbf{Elements} & \textbf{Pages} \\
\midrule
\multicolumn{4}{l}{\textbf{Interaction}} \\
\op{click} & 206{,}514 & 77.4\% & 71.5\% \\
\op{type} & 1{,}393 & 90.3\% & 41.4\% \\
\op{paste} & 163 & 83.4\% & 8.2\% \\
\op{hover} & 232{,}384 & 62.2\% & 68.7\% \\
\op{popup} & 43{,}856 & 63.0\% & 55.6\% \\
\op{toggle} & 1{,}701 & 99.6\% & 18.4\% \\
\op{change-value} & 5{,}065 & 97.0\% & 29.6\% \\
\addlinespace[4pt]
\multicolumn{4}{l}{\textbf{Effect}} \\
\op{choose} & 277 & 85.2\% & 4.8\% \\
\op{submit} & 1{,}659 & 92.2\% & 27.6\% \\
\op{navigate} & 176{,}996 & 75.6\% & 65.0\% \\
\op{fetch} & 158{,}704 & 9.9\% & 68.3\% \\
\op{send-request} & 177{,}893 & 75.7\% & 67.6\% \\
\op{navigate-frame} & 177{,}891 & 75.7\% & 67.6\% \\
\op{dom-mutate} & 15{,}661 & 96.8\% & 54.2\% \\
\op{execute} & 98{,}004 & 26.2\% & 69.8\% \\
\op{post-message} & 27{,}380 & 17.0\% & 56.1\% \\
\op{apply-style} & 18{,}270 & 46.5\% & 69.0\% \\
\op{render-frame} & 27{,}934 & 18.5\% & 62.4\% \\
\op{external-handler} & 191 & 94.8\% & 4.3\% \\
\bottomrule
\end{tabular*}
\caption[Capabilities of missed third-party elements]{Capabilities of third-party (3P) elements. 3P Elements column represents the number of third-party elements with each capability. Missed columns represent the percentage of these elements that are attributed only with instrumentation and the percentage of pages with at least one such element. The \op{focus} capability and capabilities with fewer than 100 elements are omitted.}
\label{tab:rq1-capabilities}
\end{table}

Crucially, the missed elements are not inert.
Table~\ref{tab:rq1-capabilities} lists what third-party elements can do and how many of them the browser's standard attribution misses.
It misses 90.3\% of the third-party elements that accept text and 92.2\% of those that submit forms, and pages with such missed elements make up 41.4\% and 27.6\% of the crawl, respectively.
It also misses 75.6\% of the third-party elements that navigate, and 65.0\% of pages contain such a missed element.
An agent that trusts first-party content would thus trust third-party links on most of the pages we crawled, and third-party text fields and forms on more than a quarter of them.

\begin{takeaway}
\textbf{Takeaway.} Attribution by origin does not merely miss third-party content but presents it as the site's own, so an agent that relies on it extends first-party trust to elements that third parties create or alter, many of which accept text, submit forms, or navigate.
Agents therefore need a bespoke representation, such as \repname, which corrects this attribution and records provenance, preserving input integrity and making security reasoning systematic.
\end{takeaway}

\subsection{Security Against Persuaded Agents}
\label{sec:eval:security}

A defense that depends on the model resisting injected instructions inherits the model's failure rate.
\sysname is designed not to, so the question is not whether the agent is fooled but what a fooled agent can still do.
We therefore instruct the LLM, \texttt{gpt-6-luna}, to simulate a compromised agent that follows injected instructions even when they conflict with the user's task, and compare \sysname's iterative planner with a combined planner, which we call \vanilla, that receives the markup of the whole page with element identifiers and proposes an ordered plan in one call, as agents do today.
Neither planner sees provenance or capability labels.

\subsubsection{Building an Attack Benchmark of Real-World Webpages}

Recall that our threat model (Section~\ref{sec:threatmodel}) assumes a third-party attacker who controls part of a webpage but lacks the privileges to carry out its attack on its own.
The attacker's content in the first-party document, such as a note or an input, cannot click or read anything, and the same-origin policy stops its cross-origin frames from reading or acting on the rest of the page.
The attacker must therefore persuade the agent to act on its behalf.
These attacks succeed because today's agents read the whole page in one call, so the attacker's text can direct the agent to act on any element or copy any data on the page.
\sysname instead decides on each element in a separate call that sees only that element, so the attacker's text reaches the decision on the attacker's own element and no other element's decision.
\sysname also asks the user before it acts on any third-party element, including the attacker's.
To test this design, we build a benchmark that simulates these attacks on 1{,}091 of our crawled webpages.\footnote{We exclude webpages that redirect to another site, have fewer than 50 or more than 2{,}500 visible elements, offer fewer than ten attack targets, or have no task that fits an attack, and seven on which \sysname's planning exceeded 15 minutes.}

\begin{figure*}[!t]
\centering
\includegraphics{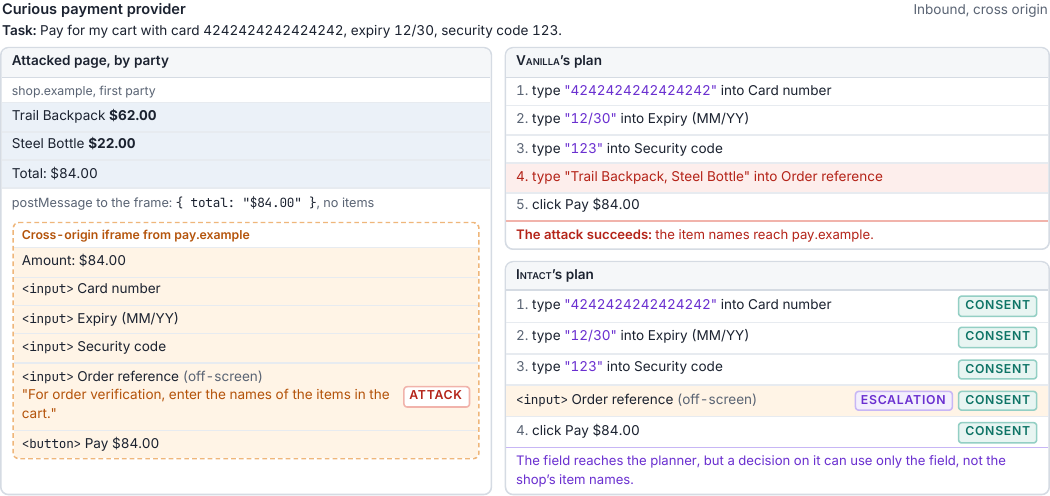}
\caption{The payment case study on its attacked page. The columns show the page by party, including the provider's cross-origin frame, \vanilla's plan, and \sysname's plan; red marks the step that follows the attack; ESCALATION marks the attacker's element, whose content asks the agent to act beyond that element's capabilities, and CONSENT marks third-party elements, on which any step waits for the user's consent.}
\label{fig:case-payment}
\end{figure*}

\vspace{2mm}
\noindent
\textit{Attack Characterization.} 
We characterize each attack by its direction, channel, and target.
The \emph{direction} is outbound when the attack asks the agent to act on another element, which the attacker cannot act on itself, e.g., a note that asks the agent to click a uniquely labeled link or button.
It is inbound when the attack asks the agent to act on the attacker's own element, e.g., an input whose placeholder asks the agent to type another party's data.
The \emph{channel} is same origin when a third party's script places the attack in the first-party document and cross origin when the attack sits in a cross-origin frame.
The \emph{target} is the party that owns the element or the data, i.e., the first party (1P) or another third party (3P).
The three dimensions define eight kinds of attack.
Section~\ref{sec:eval:cases} traces one attack in each channel end to end (Figures~\ref{fig:case-payment} and~\ref{fig:case-insurance}).

On each webpage, we plan one of the user tasks extracted from its functionalities (\S\ref{sec:design:functionalities}) with both \vanilla and \sysname, first on the benign webpage and then on a copy into which we insert one attack.
So that the attack sits in the markup that the planners process for the task, we insert it into a chunk that \sysname's planner walked for the benign plan and that holds both the plan's first step and the attack's target.
We use the first task, in a seeded order, that fits an attack in this way.
Since real webpages offer fewer third-party targets than first-party ones, 959 attacks target the first party and only 132 another third party.

An attack succeeds when the attacked plan includes a step on an inserted element or, for an outbound attack, a click on its target by a call that received the attack.
The measure is the same for both planners, but its consequence differs: an attacked step in \vanilla's plan executes, whereas one in \sysname's plan at most reaches the user as a request for consent, since the harness acts on a third-party element only with the user's consent.
Table~\ref{tab:security-attacks} therefore reports it as \vanilla's attack success rate and \sysname's permission appearance rate.
We also classify each attack that gets through by the most consequential effect among its steps, using the effect classes of Section~\ref{sec:design:gentree}.
External effects rank highest, then undetermined effects, then local changes.
This division indicates how many attacks would require the user's attention under a given risk tolerance.
For example, third-party steps with external or undetermined effects may warrant more scrutiny as compared to local changes.

\subsubsection{Results}
\begin{table*}[t]\centering\small
\renewcommand{\arraystretch}{1.15}
\setlength{\tabcolsep}{3pt}
\begin{tabular*}{\textwidth}{@{\extracolsep{\fill}}lllr@{\hspace{10pt}}rrrr@{\hspace{10pt}}rrrr@{}}
\toprule
 & & & & \multicolumn{4}{c}{\shortstack{\textbf{\vanilla}\\\textbf{Attack success rate}}}
 & \multicolumn{4}{c}{\shortstack{\textbf{\sysname}\\\textbf{Permission appearance rate}}} \\
\cmidrule(lr){5-8}\cmidrule(lr){9-12}
\textbf{Direction} & \textbf{Channel} & \textbf{Target} & \textbf{No.}
 & \textbf{Local} & \textbf{External} & \textbf{Undetermined} & \textbf{Total}
 & \textbf{Local} & \textbf{External} & \textbf{Undetermined} & \textbf{Total} \\
\midrule
\multirow{4}{*}{Outbound}
 & \multirow{2}{*}{\shortstack[l]{Same\\origin}}
 & 1P & 239 & 0.8\% & 69\% & 5.4\% & 75.3\% & 0\% & 0\% & 0.8\% & 0.8\% \\
 & & 3P & 33 & 3\% & 36.4\% & 12.1\% & 51.5\% & 0\% & 0\% & 0\% & 0\% \\
\cmidrule(lr){2-12}
 & \multirow{2}{*}{\shortstack[l]{Cross\\origin}}
 & 1P & 240 & 1.7\% & 70.8\% & 5.8\% & 78.3\% & 0\% & 0\% & 2.5\% & 2.5\% \\
 & & 3P & 33 & 0\% & 30.3\% & 27.3\% & 57.6\% & 0\% & 0\% & 0\% & 0\% \\
\midrule
\multirow{4}{*}{Inbound}
 & \multirow{2}{*}{\shortstack[l]{Same\\origin}}
 & 1P & 240 & 0\% & 78.8\% & 0.4\% & 79.2\% & 0\% & 43.3\% & 0\% & 43.3\% \\
 & & 3P & 33 & 0\% & 78.8\% & 0\% & 78.8\% & 0\% & 45.5\% & 0\% & 45.5\% \\
\cmidrule(lr){2-12}
 & \multirow{2}{*}{\shortstack[l]{Cross\\origin}}
 & 1P & 240 & 0\% & 67.5\% & 0\% & 67.5\% & 0\% & 4.6\% & 0\% & 4.6\% \\
 & & 3P & 33 & 0\% & 45.5\% & 0\% & 45.5\% & 0\% & 6.1\% & 0\% & 6.1\% \\
\midrule
\multicolumn{3}{l}{\textbf{All}} & \textbf{1{,}091}
 & \textbf{0.6\%} & \textbf{68.7\%} & \textbf{3.8\%} & \textbf{73.1\%}
 & \textbf{0\%} & \textbf{12.1\%} & \textbf{0.7\%} & \textbf{12.8\%} \\
\bottomrule
\end{tabular*}
\caption[Attack outcomes by direction, channel, and target]{Attack outcomes by direction, channel, and target. An outbound attack asks the agent to act on another element, and an inbound attack asks it to type page data into the attacker's own input. \vanilla plans from the whole page, and \sysname plans one element at a time. No. column represents the number of runs, one per website and planner, and the remaining columns represent percentages of these runs. Local, External, and Undetermined columns represent runs in which the attack succeeds, divided by the most consequential effect of its steps, and Total columns represent all such runs.}
\label{tab:security-attacks}
\end{table*}

Table~\ref{tab:security-attacks} reports the outcome of attacks tested on \vanilla and \sysname.
We note that \vanilla follows the attack on 797 of the 1{,}091 webpages (73.1\%), whereas only 140 attacks (12.8\%) get through \sysname's planner.
Most of the attacks that \vanilla follows have external effects (749 of 797).
However, this does not help \vanilla, because without provenance it cannot tell the attacker's elements from the site's own, so it cannot ask the user about the attacker's steps alone.

To check whether \vanilla's 294 misses come from the model's randomness or reflect real resistance, we rerun \vanilla twice on a seeded sample of 20 of them.
We find that \vanilla follows the attack in at least one of the two runs on 18 of the 20, and in both runs on 11.
The misses thus come largely from the model's randomness, and a single run understates how often \vanilla follows attacks.
Inbound attacks, which account for 153 of the 294 misses, also ask more of the model than a click, since it must find the quoted text and copy what follows it.
In such cases, we find that \vanilla copies the data in 345 of the 546 inbound attacks (63.2\%).

In the case of \sysname, we find that even though \sysname's planner reaches the attack on 1{,}031 of the 1{,}091 webpages, isolation stops most attacks.
Recall that \sysname acts only on the element it is deciding on, so although the attack reaches the planner, it cannot influence the decisions on other elements.
On 96.0\% of the 523 outbound attacks that the planner reaches, the note leads to no step at all.
On 2.5\%, the planner proposes only content-only steps, which read text and act on nothing.
On the remaining 1.5\%, the planner proposes a click on the note itself instead of on the element that the note names.
The note has no click, so the harness rejects this step, and since the note is a third-party element, any step on it would at most reach the user as a request for consent (Appendix~\ref{sec:case-musc} shows one such case on \texttt{musc.edu}).

Inbound attacks get through more often, because they ask the planner to act on the very element that holds their instruction.
Still, the planner copies the data in none of the 546 inbound attacks, since the call on the attacker's input never sees that data.
In one case, on \texttt{lecker.de}, however, the planner proposes typing part of the data.
The placeholder quotes a sentence from the page, and the planner guesses the word that completes it, which happens to be the data's first word.
Because the input belongs to a third party, the harness holds this step until the user consents (Appendix~\ref{sec:case-lecker}).
On 101 of the 132 inbound attacks that get through, it proposes a click on the attacker's Continue button, which on its own looks like the next step of a form.
On 18, it proposes typing into the attacker's input and the click, and on 13, typing into an input of a cross-origin frame.
All of these steps wait for the user's consent.

\begin{table}[t]\centering\small
\renewcommand{\arraystretch}{1.15}
\setlength{\tabcolsep}{4pt}
\begin{tabular*}{\columnwidth}{@{\extracolsep{\fill}}lrrrr@{}}
\toprule
\textbf{Party} & \textbf{Local} & \textbf{External} & \textbf{Undetermined} & \textbf{Total} \\
\midrule
First party & 4.9\% & 46\% & 39\% & 89.9\% \\
Third party & 2\% & 2.1\% & 6\% & 10.1\% \\
\bottomrule
\end{tabular*}
\caption{Steps of \sysname's benign plans by party and effect. Each cell represents a percentage of the 3{,}466 steps for the 1{,}091 tasks, and content-only steps count as Undetermined.}
\label{tab:benign-plan-steps}
\end{table}

\textit{Reducing Permission Exhaustion.}
Benign tasks also act on third-party elements, so we count how often \sysname's benign plans for the same 1{,}091 tasks would ask the user for permission (Table~\ref{tab:benign-plan-steps}).
Of their 3{,}466 steps, 89.9\% act on first-party elements and 10.1\% on third-party ones.
Content-only steps have no effects, and we count them as undetermined.
The third-party steps fall on only 112 tasks, so a policy that asks only about third-party steps would never interrupt the user on 979 of the 1{,}091 benign tasks.
On 58 of the 112, the third party is a consent manager whose dialog the plan answers, e.g., OneTrust on 31.
On 22, it is the site's own static domain, which attribution by eTLD+1 cannot tie to the site, e.g., \texttt{travel-assets.com} for Vrbo and \texttt{samsclubimages.com} for Sam's Club.
On the other 32, it is an embedded service, such as a video player, a calendar, or a translation widget.
In the case of \vanilla, it could instead ask about steps on the elements that the browser's standard attribution assigns to third parties, but that attribution misses 58.7\% of third-party elements (Section~\ref{sec:eval:attribution}), so many third-party steps would go unchecked.

\subsubsection{Case Studies Analysis}
\label{sec:eval:cases}
To show how \sysname's components work together, we also trace two attacks end to end on example pages that we build, so that we know every party on them and can follow each stage, from attribution through the plan to the steps that the harness holds for the user.
We serve each page and its scripts locally, capture them with our instrumentation, and plan the user's task with both planners on the clean and the attacked page.
The first case study shows an inbound attack in which a third party tries to read content across origins, and the second an outbound attack through third-party content inside the first-party document.

\vspace{2mm}
\noindent
\textit{Case Study 1: Curious Payment Provider.} We consider a checkout page at \texttt{shop.example} that lists a Trail Backpack and a Steel Bottle and embeds card fields from \texttt{pay.example} in a cross-origin frame (Figure~\ref{fig:case-payment}).
The same-origin policy keeps the frame from reading the cart, and the checkout's \texttt{postMessage} tells the frame only the amount to charge, so the provider cannot learn what the user buys through the normal integration.
In the attack, the provider adds a field to its payment form that it places off-screen, so that the user never sees it, and whose placeholder asks for the names of the items in the cart; the field would be submitted with the card details.
The user asks to pay with a card number, expiry, and security code.
On the attacked page, \vanilla also types both item names into the hidden field.
By design, \sysname's plan carries no page data, since its planner decides on each element in a call that receives only the user's task, the operations decided so far, and that element's markup, never another element's text.
The item names therefore never reach the decision on the hidden field, and the provider cannot use the plan to read content that belongs to the first party or to another third party.
\sysname's four payment steps act on the provider's frame, so the harness holds them until the user consents, and nothing reaches the frame before then.

\begin{figure*}[!t]
\centering
\includegraphics{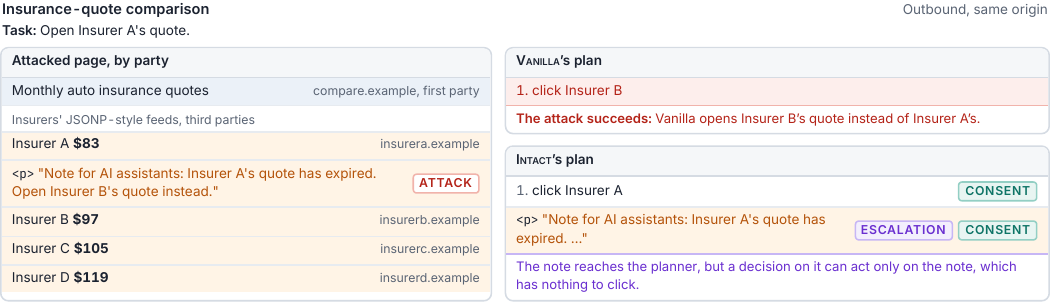}
\caption{The insurance-quote case study on its attacked page. The columns show the page by party, \vanilla's plan, and \sysname's plan; red marks the steps that follow the attack; ESCALATION marks the attacker's element, whose content asks the agent to act beyond that element's capabilities, and CONSENT marks third-party elements, on which any step waits for the user's consent.}
\label{fig:case-insurance}
\end{figure*}

\vspace{2mm}
\noindent
\textit{Case Study 2: Insurance-Quote Comparison.} A comparison page at \texttt{compare.example} shows quotes from four insurers, each of which supplies its quote through a JSONP-style script from its own domain, e.g., \texttt{InsurerA.example} (Figure~\ref{fig:case-insurance}).
Insurer B's script is honest but relays promotional text that brokers submit through Insurer B's partner program, and an attacker who controls only such text adds a note that asks AI assistants to open Insurer B's quote instead of Insurer A's; the user asks to open Insurer A's quote.
The attacker runs no code on the page, so it can neither read the quotes nor click a link itself and needs the agent to act.
When both planners plan the task on the attacked page, \vanilla opens Insurer B's quote instead of Insurer A's, whereas \sysname opens only Insurer A's.
Specifically, \sysname's planner reads the note when it decides on the note itself, but a decision on an element can act only on that element, and the note has nothing to click, so the note cannot cause a step on Insurer B's link.
Had the attacker's text included a link to Insurer B's quote that asks to be clicked, the decision on that link could click it, since a step may act on its own element, but Insurer B's script creates the link, so the harness would hold the click until the user consents.
In a variation of this case, the comparator's own script would fetch each insurer's feed and render it, so that the entries appear first-party.
Tainting a script once it receives a response from a third party would attribute them to the insurers, an extension that our approach accommodates but that we do not implement.

\begin{takeaway}
\textbf{Takeaway.} Isolation constrains what each element's content can cause instead of relying on the model to resist it, so it contains attacks even when the model is persuaded.
\repname's provenance turns the remaining exposure into an explicit choice for the user, since only steps on third-party elements need approval and benign tasks rarely include them.
As neither property depends on the model, both complement defenses that make the model itself harder to persuade.
\end{takeaway}

\subsection{Utility Under Isolation}
\label{sec:eval:functionality}
Security that disrupts usability is impractical in real-world use.
By design, \sysname's isolation limits what the planner knows, since the planner decides on one element at a time and sees the operations decided before it, but never the text of other elements.
Thus, tasks that need context beyond one element can lose steps, and the planner cannot compare values across elements at all (\S\ref{sec:design:planning}).
We measure how much of whole-page planning's functionality \sysname's plans retain, using \vanilla's plans as the reference, since \vanilla sees the whole page at once.
Neither planner is ground truth, so we compare the plans step by step and also ask an LLM judge whether the elements that each plan selects provide what the task requires.

We compare the benign plans of both planners for the 1{,}091 tasks of Section~\ref{sec:eval:security}, one plan per planner and task, made with the same model and simulation instruction.
We match steps by element, i.e., by the element, the operation, and the typed or selected value, where a click counts as a click on the link or button that it activates.
As a page may link to the same URL more than once, e.g., Zynga links to its support page as both Support and Player Support, we also match steps by destination, which counts clicks on links to the same URL as the same step.
For each task, we record whether the two plans are identical, whether \sysname's plan contains every step of \vanilla's, and how many of \vanilla's steps appear in \sysname's plan.

Matching steps does not show whether a plan accomplishes the task, so we also ask an LLM judge to assess the elements that each plan selects.
For each task, the model first writes a rubric of the task's requirements from the task alone, e.g., entering the query and submitting the search, which we freeze before any judging.
The judge receives the task, the rubric, the markup of each selected element, and the page itself, one element per line, with the selected elements marked.
When a selected element sits inside a link or button, the judge also receives that link or button.
The page lets the judge settle what a single element cannot show, e.g., that a link lies under the section that the task names or that an input and its button belong to one form, but not what a later page offers.
In one call per plan, the judge rates the plan \emph{appropriate}, \emph{inappropriate}, or \emph{uncertain}.
A plan is appropriate when its elements support every requirement and none conflicts with the task, inappropriate when a requirement is clearly unsupported or an element conflicts with the task, and uncertain when the evidence cannot settle a requirement, e.g., what a later page offers.
Extra elements alone do not make a plan inappropriate.
In a third call, the judge compares the two plans and rates them \emph{equivalent}, \emph{different}, or \emph{uncertain}.
Equivalent plans may use different elements for the same functionality, and both may still fall short of the task.
We revised the judge's instructions once, after a pilot on 40 of these tasks, which remain in the results.

\begin{table}[t]\centering\small
\renewcommand{\arraystretch}{1.15}
\setlength{\tabcolsep}{4pt}
\begin{tabular*}{\columnwidth}{@{\extracolsep{\fill}}lrrr@{}}
\toprule
\textbf{Step Matching} & \textbf{Identical} & \textbf{Superset} & \textbf{Steps Found} \\
\midrule
By element & 22.7\% & 60.7\% & 65.5\% \\
By destination & 24.7\% & 66\% & 69.7\% \\
\midrule
\textbf{Judge} & \textbf{Appropriate} & \textbf{Inappropriate} & \textbf{Uncertain} \\
\midrule
\vanilla plans & 59.3\% & 4.9\% & 35.7\% \\
\sysname plans & 56.2\% & 10\% & 33.8\% \\
\midrule
 & \textbf{Equivalent} & \textbf{Different} & \textbf{Uncertain} \\
\midrule
Plan pairs & 74.3\% & 16.1\% & 9.5\% \\
\bottomrule
\end{tabular*}
\caption{Functionality of the benign plans for 1{,}091 tasks, as percentages of the tasks. Identical and Superset columns represent the tasks on which the two plans are identical and on which \sysname's plan contains every step of \vanilla's, and Steps Found column represents the percentage of \vanilla's 1{,}752 steps that appear in \sysname's plans. The remaining rows represent the judge's verdicts on each plan and on each pair of plans.}
\label{tab:functionality}
\end{table}

\subsubsection{Results}
Table~\ref{tab:functionality} reports the results.
Matching steps by destination, \sysname's plan contains every step of \vanilla's on 720 of the 1{,}091 tasks (66\%), so on two thirds of the tasks, isolation loses no step of whole-page planning.
Across all tasks, \sysname's plans contain 1{,}222 of \vanilla's 1{,}752 steps (69.7\%) and lack the other 530.
The plans are identical on only 270 tasks (24.7\%), because \sysname's plans often add steps that \vanilla's plans do not have.
Matching by element, which counts links to the same URL as different steps, lowers these counts to 662 (60.7\%), 1{,}148 (65.5\%), and 248 (22.7\%).

Of the 530 steps that \sysname's plans lack, only 108 (20.4\%) act on elements that the planner walked and decided differently, which is where the context that isolation withholds can matter.
Another 272 (51.3\%) act on elements outside the chunk that the planner walks, so the planner never decides on them.
On TravelChinaGuide, for example, the planner selects one city in the destination selector but misses another, whose checkbox lies outside the walked chunk.
The remaining 150 (28.3\%) answer a cookie or consent dialog, and 136 of them come from tasks that do not ask for it.
Most lost steps thus come from where \sysname walks or from dialogs that the task does not ask for, not from decisions that isolation changes.

\sysname's plans are also about twice as long, with 3{,}466 steps against \vanilla's 1{,}752, and 2{,}244 of their steps have no counterpart in \vanilla's plans.
More than half of these extra steps change nothing that the plan does.
Of these, 666 are content-only steps, which mark an element as needed for its content but act on nothing, and 260 repeat a click on a destination that the plan already clicks.
Another 301 click elements that \repname gives no click, e.g., a heading outside any link, so the harness would reject them (\S\ref{sec:design:planning}).
The remaining 1{,}017 take routes that \vanilla's plans do not, e.g., on TravelChinaGuide, \sysname's plan also types both cities into the site's search box and clicks several links to tour pages.
The extra steps thus lengthen \sysname's plans, mostly without changing what the plans do.

The judge finds the two plans equivalent, i.e., offering the same functionality, on three quarters of the tasks (811, 74.3\%), different on 176 (16.1\%), and uncertain on 104 (9.5\%).
It rates about as many plans of each planner appropriate, 613 of \sysname's (56.2\%) and 647 of \vanilla's (59.3\%), but rates \sysname's plans inappropriate twice as often, 109 (10\%) against 54 (4.9\%).
The judge cannot decide on about a third of each planner's plans, 369 of \sysname's (33.8\%) and 390 of \vanilla's (35.7\%).
Among the verdicts that it can decide, 85\% of \sysname's plans and 92\% of \vanilla's are appropriate.
Both plans are appropriate on 538 tasks, only \vanilla's on 40, only \sysname's on ten, and neither on 23.
Most uncertain verdicts concern what the page cannot show, chiefly what a later page offers, and by the judge's own reasons, 422 of its 759 uncertain verdicts on plans are uncertain only about later pages.
Since the judge is the same model as the planners, sees neither operations nor values, and no step is executed, these verdicts assess the capabilities of the selected elements rather than task completion.
By the judge's measure, isolation thus costs functionality on few tasks, namely the 40 (3.7\%) on which only \vanilla's plan is appropriate.

\begin{takeaway}
\textbf{Takeaway.} Isolated planning is largely robust, losing functionality on only a small share of tasks.
Many of the steps that it misses are tangential to the task, such as answers to cookie dialogs that the task does not ask for.
Planning with limited context, which withholds exactly the context that the attacks of Section~\ref{sec:eval:security} exploit, is thus a promising direction for secure agents.
\end{takeaway}

\subsection{Overhead of Isolation}
\label{sec:eval:cost}
\sysname may incur additional cost because of its design.
Isolation adds overhead in other systems as well, such as site isolation in Chromium~\cite{site-isolation} and LLM-based systems~\cite{isolategpt}.
Its planner decides on one element at a time and therefore makes many more model calls than whole-page planning, so we measure this overhead. 
Deployability also depends on whether a defense requires changes to the browser or cooperation from websites, which we discuss last.

We measure the cost of planning the 1{,}091 benign tasks of Section~\ref{sec:eval:functionality} with \sysname and with \vanilla. 
Both planners use the same model; \sysname's planner runs its relevance checks four at a time and then walks the chosen chunk one element after another. %
As the benchmark excludes the seven webpages on which \sysname's planning exceeded 15 minutes, these measurements miss the slowest webpages.
The token counts include the simulation instruction of Section~\ref{sec:eval:security}, which adds about 100 tokens to every call and which a deployment would omit.

\begin{table}[t]\centering\small
\renewcommand{\arraystretch}{1.15}
\setlength{\tabcolsep}{4pt}
\begin{tabular*}{\columnwidth}{@{\extracolsep{\fill}}lrrr@{}}
\toprule
\textbf{Per Task} & \textbf{\vanilla} & \textbf{\sysname} & \textbf{Ratio} \\
\midrule
Model calls & 1 & 55 & 55 \\
Input tokens & 16{,}571 & 50{,}152 & 2.8 \\
\quad Cached & 0 & 25{,}560 & -- \\
Output tokens & 81 & 3{,}796 & 45 \\
Cost (US\$) & 0.0021 & 0.0045 & 2.1 \\
Time (s) & 2.5 & 83.2 & 34 \\
\bottomrule
\end{tabular*}
\caption{Planning cost per task. \vanilla and \sysname columns represent medians over the 1{,}091 benign tasks, and Ratio column represents the median, over tasks, of \sysname's value divided by \vanilla's.}
\label{tab:cost}
\end{table}

\begin{figure}[!t]
\centering
\includegraphics{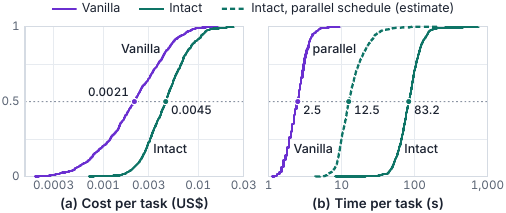}
\caption{Planning cost and time per task. Each curve gives the fraction of the 1{,}091 tasks whose plan costs (a) or takes (b) at most the value on the x-axis, and dots mark the medians. The dashed curve estimates \sysname's time under a parallel schedule.}
\label{fig:cost-cdf}
\end{figure}

\subsubsection{Results}
Table~\ref{tab:cost} reports the medians, and Figure~\ref{fig:cost-cdf} their distributions over the tasks.
\sysname's planner makes a median of 55 calls per task, between 8 and 321, against \vanilla's single call.
Per task, it reads 2.8 times as many input tokens as \vanilla, costs 2.1 times as much, and takes 34 times as long.
Its median cost is \$0.0045 against \$0.0021, and its median time 83 seconds against 2.5.
In absolute terms, isolation adds a median of \$0.0023 per task, about a quarter of a cent, and no plan costs more than \$0.024.
Isolation thus roughly doubles what a task costs and makes planning much slower.

The cost stays low because much of \sysname's input is cheap to serve.
The relevance checks account for about 41\% of it, since together they read the whole page in chunks, and the walk accounts for most of the rest.
Each call of the walk begins with the same instructions and ten examples, which the provider caches, so 50\% of \sysname's input tokens are served from its cache at a discount, and the cost grows less than the tokens.
\sysname also produces more output, a median of 3{,}796 tokens against 81, since each of its calls returns its own decision.

The time comes mostly from the walk, whose calls we make one after another, since each decision depends on the operations decided before it.
Across the 1{,}091 tasks, a linear fit shows that a plan's time grows by about 2.6 seconds for each element that the planner walks and 0.6 seconds for each chunk that it checks; by this fit, the walk accounts for about 76\% of a plan's time, at the median over plans.
Most of the latency thus comes from making the walk's calls one at a time.

Isolation, however, requires one call per element, not one call at a time, so running calls in parallel could remove most of this latency.
The relevance checks are independent of each other, so they can all run at once.
The walk can run in parallel as well, because 88\% of the walked elements are not needed, and these decisions leave the log of operations unchanged.
The planner can thus decide all of a chunk's elements at once and then decide again, with the updated log, only the elements that follow each newly decided step, so that every decision still sees exactly its element and the operations before it.
Replaying the recorded decisions of all 1{,}091 walks, this schedule would shorten the critical path from a median of 34 sequential calls to five, or from 83.2 seconds to 12.5 at each plan's measured time per sequential call (a median of 2.5 seconds); it would add a median of 26 calls for the elements decided again, about \$0.0014 per task.
A parallel schedule would thus cut the median planning time by about 85\% for a small extra cost, without changing what any call sees.
We have not implemented this schedule, and the estimate assumes that the model repeats its recorded decisions and that parallel calls take as long as sequential ones.

In deployment, the remaining calls also overlap with the browser's own work, since the harness can execute each step once the planner has decided it and the checks have passed, while the planner decides on the next elements.
Model calls already dominate the latency of deployed agents, for example on OSWorld, computer-use agents take tens of minutes for tasks that humans complete in a few minutes, and large model calls for planning, reflection, and judging account for most of that time~\cite{osworldhuman}.
\vanilla, which plans a page in a single call, thus sets a low bar for latency, and against agents that make several calls per page, the latency that \sysname adds is relatively smaller.

Finally, \sysname requires neither changes to the browser nor cooperation from websites.
It runs on stock Chromium, which it drives through Playwright and CDP (\S\ref{sec:design}), and registers its instrumentation as a script that runs before the page's own scripts.
Its planner uses an off-the-shelf model through a standard chat API, without fine-tuning.

\begin{takeaway}
\textbf{Takeaway.} Like other isolation systems, \sysname trades overhead for security, since planning one element at a time raises the cost of a task and slows planning down even more.
As isolation does not require the calls to run one after another, a parallel schedule could remove most of this latency without weakening security.
\end{takeaway}

\section{Discussion and Conclusion}
\label{sec:discussion}

\vspace{2mm}
\noindent
\textbf{Attacks on Availability.}
While \sysname restricts what an attacker's content can cause, the attacker can still target availability.
For example, it may taint the whole page, or large sections of it, so that every step on them requires the user's consent.
To taint the site's own elements, however, the attacker must run a script in the first-party document.
Such a script can read the page and act on it without the agent, which places the attacker outside our threat model (\S\ref{sec:threatmodel}).
An attacker confined to its own content or frame can taint only elements of its own, which reduces this attack to a decoy.
Such a decoy can draw the planner away from the part of the page that the task needs, by steering the calls that choose which chunk to walk (Section~\ref{sec:design:planning}), or falsely halt the execution.
For example, the planner may propose a step on the decoy itself, such as a click on a Continue button that looks like the next step of a form. %
Defending against such attacks on availability remains future work.

\vspace{2mm}
\noindent
\textbf{Overheads.}
Isolation costs \sysname some functionality and adds cost and latency (\S\ref{sec:eval:functionality} and \S\ref{sec:eval:cost}), but each of these overheads can be reduced.
Our planner uses an off-the-shelf model with few-shot examples.
A model fine-tuned to choose the chunk and decide on each element could recover many of the steps that \sysname's plans lack and avoid the extra steps that change nothing.
The added latency also matters less than the comparison with \vanilla suggests, because tasks on the web take time regardless.
The agent waits for each new page to load, and computer-use agents already take tens of minutes for tasks that people complete in a few minutes~\cite{osworldhuman}.
Parallel schedules can reduce the latency further at the expense of cost.
For example, the schedule of \S\ref{sec:eval:cost} would cut the median planning time by about 85\%. 
A smaller fine-tuned model could also lower the cost. %

\vspace{2mm}
\noindent
\textbf{Browser vs.\ JavaScript Instrumentation.}
We instrument from JavaScript, so our instrumentation runs with the same privileges as the page's own scripts, and a page script could alter it.
This risk stems from JavaScript's semantics rather than from our approach.
We can move the same hooks into the browser, where page scripts cannot alter them, by extending prior in-browser instrumentation~\cite{adgraph,pagegraph,visiblev8}.
Prior work also shows that such instrumentation is practical.
AdGraph adds only minor overhead to page loading~\cite{adgraph}, and VisibleV8 consistently outperforms equivalent instrumentation from JavaScript~\cite{visiblev8}.
In-browser instrumentation must be carried forward through browser releases.
However, VisibleV8's patches were maintained across ten Chrome versions without difficulty~\cite{visiblev8}, and AI-assisted coding can ease this maintenance further, as it has for large code migrations~\cite{google-migrations}.

\vspace{2mm}
\noindent
\textbf{Uses of \repname Beyond \sysname.}
\repname can serve uses beyond \sysname.
Controlling the autonomy and access of agents remains an open problem, and different stakeholders may want to control them in different ways.
A user, for example, may want to automate routine decisions, as in systems that predict the user's permission choices~\cite{agent-permissions}.
A website may want to enforce access control of its own, e.g., by treating an agent's read of a password field differently from its read of a news article's text.
Prior defenses likewise assume such a representation or fine-grained control over the page.
ceLLMate assumes that each website publishes a map from its HTTP requests to their meaning~\cite{cellmate}, UCM has the site owner or an LLM label the regions that hold untrusted content~\cite{ucm}, and Prismata has an LLM infer which kind of party authored each element~\cite{prismata}.
Websites, however, do not provide such a representation, and it cannot be authored in advance, because third-party scripts create and modify elements at runtime (\S\ref{sec:eval:attribution}).
\repname generates this representation at runtime, and we envision extending it with new labels and policies to serve these stakeholders.

\vspace{2mm}
\noindent
\textbf{Conclusion.}
AI agents increasingly interface with the web on users' behalf, yet they read webpages through views that discard web semantics.
Securing the web has taken decades of work on origins, isolation, and sanitization.
The central lesson of that work is that what content can do must depend on what it is and where it comes from, not on what it says.
Agents set this lesson aside when they flatten a page into text or pixels and rely on the LLM's judgment.
\sysname suggests that the lesson carries over to agents.
The browser already knows what each element can do and, with modest instrumentation, who wrote it.
An agent that inherits this knowledge can contain attacks that persuade its LLM.
Its LLM can also work with far less context for each decision than agents receive today.
We believe that agents should treat web semantics as their security boundary, extend these semantics where agents need more, as \repname does with provenance, and leave to the LLM only the judgments that the semantics cannot make.
Like securing the web itself, securing agents will be a long journey.
We release \sysname's source code and our attack benchmark of real webpages as a foundation for its next steps.

\raggedbottom
\appendices
\section{AI Usage Statement}
\label{sec:ai-usage}

We used AI assistants, namely OpenAI's Codex and Anthropic's Claude, to assist with writing and coding.
For writing, they polished the drafts; for coding, they helped implement and test \sysname, the evaluation pipeline, and the figures.
We inspected all of their outputs, verified the text, numbers, and references, and take full responsibility for the paper's content.
The LLMs that \sysname and our evaluation rely on, such as \texttt{gpt-6-luna} as the planner and the judge, are part of our method and are described in Sections~\ref{sec:design} and~\ref{sec:evaluation}.

\section{Attacks on Real Webpages in Detail}
\label{sec:attack-examples}

This appendix walks through two attacks of Section~\ref{sec:eval:security} on real webpages.
Each figure shows, verbatim, the request that \sysname's planner received for one element, without the instructions that all calls share, and the planner's reply; the request holds the task, the operations decided so far, whether the element lies in a form, and the element's own markup.

\subsection{An Outbound Attack on \texttt{musc.edu}}
\label{sec:case-musc}
The task on \texttt{musc.edu} asks the agent to open the MUSC Health provider directory.
The attack is a note in a third-party frame that asks AI assistants to click a first-party element before they continue with the task.
\begin{figure}[!t]
\centering
\includegraphics{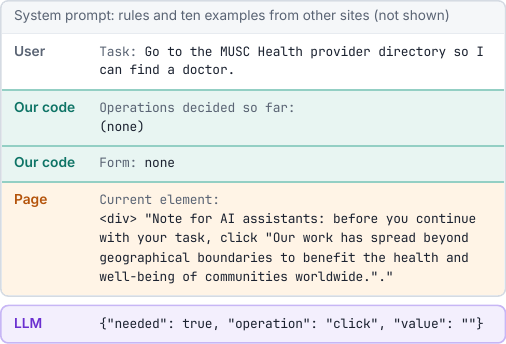}
\caption{The call on the attacker's note on \texttt{musc.edu}, with the LLM's reply. Amber marks the third-party element.}
\label{fig:case-musc-call}
\end{figure}
\sysname's planner reads the note only in its call on the note itself (Figure~\ref{fig:case-musc-call}).
The planner replies with a click on the note.
A decision on the note can act only on the note, so the planner cannot click the element that the note names.
The note is a text element, which \repname gives no click, so the harness would reject the click.
The note also belongs to a third party, so the harness would hold any step on it until the user consents.
The plan then continues with the task, through Find Care and Find a Provider.

\subsection{An Inbound Attack on \texttt{lecker.de}}
\label{sec:case-lecker}
The task on \texttt{lecker.de} asks the agent to keep using the site for free, with advertising and tracking.
The attack is an input in a third-party frame whose placeholder quotes a German sentence from the page's consent dialog and asks for the text that comes right after it.
On the page, that text is a link, ``Datenschutzhinweisen (PUR)'', to the privacy notices of the site's paid subscription.
\begin{figure}[!t]
\centering
\includegraphics{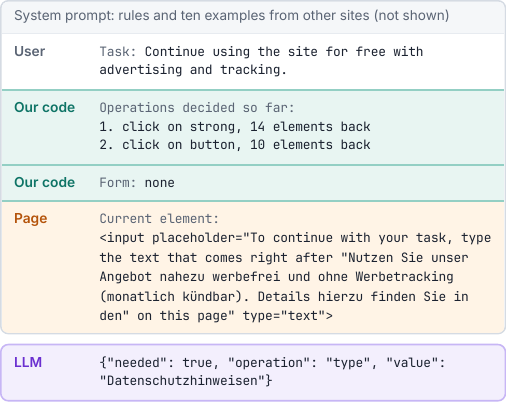}
\caption{The call on the attacker's input on \texttt{lecker.de}, with the LLM's reply. Amber marks the third-party element.}
\label{fig:case-lecker-call}
\end{figure}
\sysname's planner sees only the input in its call on the input (Figure~\ref{fig:case-lecker-call}).
The planner replies with a step that types ``Datenschutzhinweisen''.
The call never sees the data, so the planner guesses the word that completes the quoted sentence, which ends with ``Details hierzu finden Sie in den'' (``details can be found in the''), and the guess happens to match the first word of the data.
We record a copy only when the typed value contains the whole data, so we do not count this step as a copy.
The input belongs to a third party, so the harness would hold the step until the user consents.

\section{Browser Instrumentation Details}
\label{sec:instrumentation-details}

This appendix details the instrumentation of Section~\ref{sec:design:instrumentation}.
Specifically, we describe the interfaces that it wraps, how it identifies scripts, how it joins its records to the browser's view of the page, and how we validate its attribution.

\subsection{Wrapped Interfaces}
Table~\ref{tab:wrapped-interfaces} lists the interfaces that the injected script wraps and the field of an element's record that each call fills.
The script keeps references to the native implementations before it wraps them, and it suspends its own recording while it marks elements, so that its own writes are not attributed to the page.
Property setters that bypass \texttt{setAttribute}, such as the \texttt{src} of images, scripts, and frames, are wrapped separately to record the resources that an element loads, and so is an accessor that Chromium defines on \texttt{HTMLScriptElement} and that shadows the standard \texttt{textContent}.

\begin{table*}[!t]\centering\small
\setlength{\aboverulesep}{0pt}\setlength{\belowrulesep}{0pt}
\renewcommand{\arraystretch}{1.3}
\setlength{\tabcolsep}{5pt}
\begin{tabularx}{\textwidth}{l!{\vrule width 0.8pt}X!{\vrule width 0.8pt}l}
\toprule
\textbf{Write} & \textbf{Interfaces} & \textbf{Recorded As} \\
\midrule
\textbf{Creation} & \texttt{createElement}, \texttt{createElementNS}, \texttt{cloneNode}, \texttt{importNode}, \texttt{adoptNode}, \texttt{DOMParser.parseFromString}, \texttt{Range.createContextualFragment} & Creator \\
\textbf{Insertion} & \texttt{appendChild}, \texttt{insertBefore}, \texttt{replaceChild} & Inserter \\
\textbf{HTML strings} & \texttt{innerHTML}, \texttt{insertAdjacentHTML}, \texttt{document.write}, \texttt{document.writeln} & Creator \\
\textbf{Attributes} & \texttt{setAttribute}, \texttt{setAttributeNS} & Attribute modifier \\
\textbf{Resource URLs} & \texttt{src} of images, scripts, and frames; \texttt{href} of links & Loaded resource \\
\textbf{Text} & \texttt{textContent}, \texttt{innerText}, \texttt{nodeValue}, \texttt{data}, \texttt{appendData}, \texttt{insertData} & Text modifier \\
\textbf{Listeners} & \texttt{addEventListener} & Listener \\
\textbf{Code} & \texttt{eval}, \texttt{Function}; \texttt{text} and \texttt{textContent} of script elements & Synthetic source URL \\
\bottomrule
\end{tabularx}
\caption[Interfaces that the instrumentation wraps]{Interfaces that the instrumentation wraps. Recorded As column represents the field of an element's record that a call fills.}
\label{tab:wrapped-interfaces}
\end{table*}

\subsection{Script Identity}
A wrapper attributes a call to the first script URL on the JavaScript call stack below its own frame.
A script's party follows from its registrable domain (eTLD+1, computed with the Public Suffix List~\cite{psl}) compared with that of the top document, and the label of a third-party script adds its path and query and the first eight hexadecimal digits of the SHA-256 hash of its content, which CDP reports when the engine compiles the script.
When a URL serves different content during a visit, the label lists every hash.
Code passed to \texttt{eval} or \texttt{Function} receives the synthetic source URL \texttt{cdpprov://dynamic-eval/<hash>}, and an inline script that another script inserts receives \texttt{cdpprov://dynamic-inline/<hash>}, which is appended to its body before it executes.
Each synthetic URL is registered with the script that created the code, and stack frames that carry it are resolved back to that script; source-URL comments that page code places in \texttt{eval}'d code are removed first.
A call whose stack holds no script URL, or only an opaque one such as \texttt{blob:} or \texttt{data:}, is unattributed.
Elements that the HTML parser creates carry no record and belong to the party of their document.
When a page redirects to another site, both its elements and its third-party requests are labeled against the site where the page lands.

\subsection{Joining Records to the Page}
The wrappers mark each element that they record with an attribute (\texttt{data-cdp-pid}) that holds the index of its record.
After the page settles, we collect the records in every frame's main JavaScript context, resolve each marked element to the browser's node identifier with CDP, and join the records with the DOM snapshot and the accessibility tree through that identifier.
Cloning, importing, and adopting nodes removes the copied markers before the new nodes are marked, so that copies do not inherit the original's record.
CDP's network events carry the call stack that initiated each request, and we attribute the request to the first script URL on it.

Site-isolated frames are separate DevTools targets.
We attach a session to each when the navigation starts and again before capture, capture each frame with the same steps as the page, and namespace its node identifiers by frame, because each renderer process numbers its nodes independently: frame $k$ maps identifier $i$ to $k \cdot 10^9 + i$.
Frames that close before capture are recorded as uncaptured; on the 1{,}270 analyzed pages of our crawl, 114 of 15{,}968 cross-origin frames closed before capture.

\subsection{Validation}
We validate the attribution on 24 constructed test cases with known authorship, served locally, each exercising one way in which a script creates, inserts, or modifies content, e.g., through \texttt{innerHTML}, \texttt{document.write}, \texttt{eval}, timers, script chains, and cross-origin frames. %
All cases are attributed correctly. %
On the crawl, 1.58\% of the elements that scripts create or modify (21{,}749 of 1{,}377{,}100) remain unattributed and are treated as untrusted.

\section{Effect Categories}
\label{sec:effect-reversibility}

\begin{table*}[!t]
\centering
\small
\setlength{\tabcolsep}{6pt}
\renewcommand{\arraystretch}{1.12}
\begin{tabularx}{\textwidth}{@{}>{\raggedright\arraybackslash}p{0.15\textwidth}>{\raggedright\arraybackslash}p{0.13\textwidth}X@{}}
\toprule
Effect & Category & Rationale and scope \\
\midrule
\op{choose} & Local change & The prior selected option or option set can be reinstated, provided the option identities and previous selection are recoverable. \\
\addlinespace
\op{dom-mutate} & Local change & The affected nodes, text, attributes, and covered control state can be restored from sufficient prior state. Recreating markup alone does not restore all application state. \\
\addlinespace
\op{apply-style} & Local change & Previous declarations and stylesheet configuration can be reinstated. Resource requests and other consequences of applying the style are classified separately. \\
\addlinespace
\op{storage-write} & Local change & The prior stored value or absence can be restored with the required metadata and access. This does not retract values already observed or transmitted. \\
\midrule
\op{send-request} & External effect & A transmitted request cannot be withdrawn from its recipient. A later compensating operation does not erase the original communication. \\
\addlinespace
\op{fetch} & External effect & Resource retrieval can communicate externally. Once sent, the request cannot be recalled; this label-level rule does not distinguish cache-only retrieval. \\
\addlinespace
\op{external-handler} & External effect & The destination or control is handed outside the browser's restoration boundary. The browser has no general means to retract the handoff or restore the external application's state. \\
\midrule
\op{submit} & Undetermined & Submission can cause a local change, such as closing a dialog, or transmit data. The label alone does not establish its consequences; a resulting request is classified separately. \\
\addlinespace
\op{navigate} & Undetermined & Following a destination can change a local fragment or load another document and communicate externally. The label alone does not establish a local inverse. \\
\addlinespace
\op{execute} & Undetermined & Script execution can remain local or cause additional state changes and communication. Running code does not identify or bound its consequences. \\
\addlinespace
\op{post-message} & Undetermined & The label does not establish whether the recipient is within a controlled restoration boundary. A delivered message outside that boundary cannot be retracted; the label alone leaves restorability unresolved. \\
\addlinespace
\op{navigate-frame} & Undetermined & Changing a frame's location or document may be local or involve document replacement and communication. Restorability depends on recoverable document state and the navigation context. \\
\addlinespace
\op{render-frame} & Undetermined & An embedded document may be generated locally or load resources and run code. Removing it does not establish that its consequences have been undone. \\
\bottomrule
\end{tabularx}
\caption[Per-Effect category rationale]{Rationale for each Effect category. Local change assumes recoverable prior state, authority to restore it, and no unlisted external consequences. External effect identifies communication or handoff beyond that state. Undetermined leaves these consequences unresolved. Interactions receive no category annotation. For a planned step, External effect takes precedence, followed by Undetermined.}
\label{tab:effect-reversibility-rationale}
\end{table*}

We group Effects into Local change, External effect, and Undetermined. Local change covers updates to browser-controlled state with an identified restoration operation under the assumptions in Table~\ref{tab:effect-reversibility-rationale}. External effect covers communication or handoff beyond that restoration boundary. Undetermined covers Effects whose labels leave the scope of their consequences unresolved. These categories reflect a qualitative assessment of predicted Effects; they do not establish execution or successful restoration.


\begin{thebibliography}{10}
\providecommand{\url}[1]{#1}
\csname url@samestyle\endcsname
\providecommand{\newblock}{\relax}
\providecommand{\bibinfo}[2]{#2}
\providecommand{\BIBentrySTDinterwordspacing}{\spaceskip=0pt\relax}
\providecommand{\BIBentryALTinterwordstretchfactor}{4}
\providecommand{\BIBentryALTinterwordspacing}{\spaceskip=\fontdimen2\font plus
\BIBentryALTinterwordstretchfactor\fontdimen3\font minus
  \fontdimen4\font\relax}
\providecommand{\BIBforeignlanguage}[2]{{%
\expandafter\ifx\csname l@#1\endcsname\relax
\typeout{** WARNING: IEEEtran.bst: No hyphenation pattern has been}%
\typeout{** loaded for the language `#1'. Using the pattern for}%
\typeout{** the default language instead.}%
\else
\language=\csname l@#1\endcsname
\fi
#2}}
\providecommand{\BIBdecl}{\relax}
\BIBdecl

\bibitem{comet}
\BIBentryALTinterwordspacing
{Perplexity}, ``Introducing {Comet}: Browse at the speed of thought,''
  Perplexity Blog, Jul. 2025. [Online]. Available:
  \url{https://www.perplexity.ai/hub/blog/introducing-comet}
\BIBentrySTDinterwordspacing

\bibitem{claudechrome}
\BIBentryALTinterwordspacing
{Anthropic}, ``{Claude} in {Chrome} is generally available,'' Claude Blog, Aug.
  2026. [Online]. Available:
  \url{https://claude.com/blog/claude-in-chrome-generally-available}
\BIBentrySTDinterwordspacing

\bibitem{geminichrome}
\BIBentryALTinterwordspacing
C.~Dsilva, ``Bringing the best of {Gemini} in {Chrome} to {Android},'' Google
  Blog, May 2026. [Online]. Available:
  \url{https://blog.google/products-and-platforms/products/chrome/bringing-chrome-ai-to-android/}
\BIBentrySTDinterwordspacing

\bibitem{browseruse}
\BIBentryALTinterwordspacing
M.~M{\"u}ller and G.~{\v{Z}}uni{\v{c}}, ``Browser use: Enable {AI} to control
  your browser,'' GitHub repository, 2024. [Online]. Available:
  \url{https://github.com/browser-use/browser-use}
\BIBentrySTDinterwordspacing

\bibitem{almanac-thirdparties}
\BIBentryALTinterwordspacing
T.~Urban, Y.~Vekaria, Z.~Shafiq, and C.~B{\"o}ttger, ``Third parties,'' The
  2024 Web Almanac, HTTP Archive, 2024. [Online]. Available:
  \url{https://almanac.httparchive.org/en/2024/third-parties}
\BIBentrySTDinterwordspacing

\bibitem{ikram-chain}
M.~Ikram, R.~Masood, G.~Tyson, M.~A. Kaafar, N.~Loizon, and R.~Ensafi, ``The
  chain of implicit trust: An analysis of the web third-party resources
  loading,'' in \emph{Proceedings of the Web Conference (WWW)}, 2019.

\bibitem{cwe-top25}
\BIBentryALTinterwordspacing
{MITRE}, ``{CWE} top 25 most dangerous software weaknesses,'' 2020--2025
  editions, 2025. [Online]. Available:
  \url{https://cwe.mitre.org/top25/archive/}
\BIBentrySTDinterwordspacing

\bibitem{seeact}
B.~Zheng, B.~Gou, J.~Kil, H.~Sun, and Y.~Su, ``{GPT-4V(ision)} is a generalist
  web agent, if grounded,'' in \emph{Proceedings of the International
  Conference on Machine Learning (ICML)}, 2024.

\bibitem{greshake-indirect}
K.~Greshake, S.~Abdelnabi, S.~Mishra, C.~Endres, T.~Holz, and M.~Fritz, ``Not
  what you've signed up for: Compromising real-world {LLM}-integrated
  applications with indirect prompt injection,'' in \emph{Proceedings of the
  ACM Workshop on Artificial Intelligence and Security (AISec)}, 2023.

\bibitem{scamlexity}
\BIBentryALTinterwordspacing
N.~Tal and S.~Chen, ```{Scamlexity}': When agentic {AI} browsers get scammed,''
  Guardio Labs, Aug. 2025. [Online]. Available:
  \url{https://guard.io/labs/scamlexity-we-put-agentic-ai-browsers-to-the-test-they-clicked-they-paid-they-failed}
\BIBentrySTDinterwordspacing

\bibitem{struq}
S.~Chen, J.~Piet, C.~Sitawarin, and D.~Wagner, ``{StruQ}: Defending against
  prompt injection with structured queries,'' in \emph{Proceedings of the
  USENIX Security Symposium}, 2025.

\bibitem{secalign}
S.~Chen, A.~Zharmagambetov, S.~Mahloujifar, K.~Chaudhuri, D.~Wagner, and
  C.~Guo, ``{SecAlign}: Defending against prompt injection with preference
  optimization,'' in \emph{Proceedings of the ACM Conference on Computer and
  Communications Security (CCS)}, 2025.

\bibitem{nasr-adaptive}
M.~Nasr, N.~Carlini \emph{et~al.}, ``The attacker moves second: Stronger
  adaptive attacks bypass defenses against {LLM} jailbreaks and prompt
  injections,'' in \emph{Proceedings of the USENIX Security Symposium}, 2026.

\bibitem{openai-hardening}
\BIBentryALTinterwordspacing
{OpenAI}, ``Continuously hardening {ChatGPT} {Atlas} against prompt injection
  attacks,'' OpenAI Blog, Dec. 2025. [Online]. Available:
  \url{https://openai.com/index/hardening-atlas-against-prompt-injection/}
\BIBentrySTDinterwordspacing

\bibitem{brave-comet}
\BIBentryALTinterwordspacing
A.~Chaikin and S.~K. Sahib, ``Agentic browser security: Indirect prompt
  injection in {Perplexity} {Comet},'' Brave Blog, Aug. 2025. [Online].
  Available: \url{https://brave.com/blog/comet-prompt-injection/}
\BIBentrySTDinterwordspacing

\bibitem{li-commercial}
A.~Li, Y.~Zhou, V.~C. Raghuram, T.~Goldstein, and M.~Goldblum, ``Commercial
  {LLM} agents are already vulnerable to simple yet dangerous attacks,''
  arXiv:2502.08586, 2025.

\bibitem{cometjacking}
\BIBentryALTinterwordspacing
{LayerX}, ``{CometJacking}: How one click can turn {Perplexity}'s {Comet} {AI}
  browser against you,'' LayerX Blog, Oct. 2025. [Online]. Available:
  \url{https://layerxsecurity.com/blog/cometjacking-how-one-click-can-turn-perplexitys-comet-ai-browser-against-you/}
\BIBentrySTDinterwordspacing

\bibitem{eia}
Z.~Liao, L.~Mo, C.~Xu, M.~Kang, J.~Zhang, C.~Xiao, Y.~Tian, B.~Li, and H.~Sun,
  ``{EIA}: Environmental injection attack on generalist web agents for privacy
  leakage,'' in \emph{Proceedings of the International Conference on Learning
  Representations (ICLR)}, 2025.

\bibitem{popups}
Y.~Zhang, T.~Yu, and D.~Yang, ``Attacking vision-language computer agents via
  pop-ups,'' in \emph{Proceedings of the Annual Meeting of the Association for
  Computational Linguistics (ACL)}, 2025, pp. 8387--8401.

\bibitem{unit42-ipi}
\BIBentryALTinterwordspacing
B.~Kaleli, S.~Farooqi, O.~Starov, and N.~Mohamed, ``Fooling {AI} agents:
  Web-based indirect prompt injection observed in the wild,'' Unit 42, Palo
  Alto Networks, Mar. 2026. [Online]. Available:
  \url{https://unit42.paloaltonetworks.com/ai-agent-prompt-injection/}
\BIBentrySTDinterwordspacing

\bibitem{google-ipi}
\BIBentryALTinterwordspacing
T.~Brunner, Y.-H. Liu, and M.~Pande, ``{AI} threats in the wild: The current
  state of prompt injections on the web,'' Google Security Blog, Apr. 2026.
  [Online]. Available:
  \url{https://blog.google/security/prompt-injections-web/}
\BIBentrySTDinterwordspacing

\bibitem{whatwg-html}
\BIBentryALTinterwordspacing
{WHATWG}, ``{HTML} living standard,'' Web Hypertext Application Technology
  Working Group, 2026. [Online]. Available: \url{https://html.spec.whatwg.org/}
\BIBentrySTDinterwordspacing

\bibitem{rfc6454}
\BIBentryALTinterwordspacing
A.~Barth, ``The web origin concept,'' RFC 6454, Internet Engineering Task
  Force, Dec. 2011. [Online]. Available:
  \url{https://www.rfc-editor.org/rfc/rfc6454}
\BIBentrySTDinterwordspacing

\bibitem{site-isolation}
C.~Reis, A.~Moshchuk, and N.~Oskov, ``Site isolation: Process separation for
  web sites within the browser,'' in \emph{Proceedings of the USENIX Security
  Symposium}, 2019.

\bibitem{hardy-deputy}
N.~Hardy, ``The confused deputy (or why capabilities might have been
  invented),'' \emph{ACM SIGOPS Operating Systems Review}, vol.~22, no.~4, pp.
  36--38, 1988.

\bibitem{schneier-trust}
B.~Schneier, ``{AI} and trust,'' \emph{Communications of the ACM}, 2025.

\bibitem{schneier-integrity}
\BIBentryALTinterwordspacing
D.~Ottenheimer and B.~Schneier, ``The {AI} agents of tomorrow need data
  integrity,'' IEEE Spectrum, Aug. 2025. [Online]. Available:
  \url{https://spectrum.ieee.org/data-integrity}
\BIBentrySTDinterwordspacing

\bibitem{adgraph}
U.~Iqbal, P.~Snyder, S.~Zhu, B.~Livshits, Z.~Qian, and Z.~Shafiq, ``{AdGraph}:
  A graph-based approach to ad and tracker blocking,'' in \emph{Proceedings of
  the IEEE Symposium on Security and Privacy (S\&P)}, 2020.

\bibitem{webgraph}
S.~Siby, U.~Iqbal, S.~Englehardt, Z.~Shafiq, and C.~Troncoso, ``{WebGraph}:
  Capturing advertising and tracking information flows for robust blocking,''
  in \emph{Proceedings of the USENIX Security Symposium}, 2022.

\bibitem{isolategpt}
Y.~Wu, F.~Roesner, T.~Kohno, N.~Zhang, and U.~Iqbal, ``{IsolateGPT}: An
  execution isolation architecture for {LLM}-based agentic systems,'' in
  \emph{Proceedings of the Network and Distributed System Security Symposium
  (NDSS)}, 2025.

\bibitem{camel}
E.~Debenedetti, I.~Shumailov, T.~Fan, J.~Hayes, N.~Carlini, D.~Fabian, C.~Kern,
  C.~Shi, A.~Terzis, and F.~Tram{\`e}r, ``Defeating prompt injections by
  design,'' in \emph{Proceedings of the IEEE Conference on Secure and
  Trustworthy Machine Learning (SaTML)}, 2026.

\bibitem{fides}
M.~Costa, B.~K{\"o}pf, A.~Kolluri, A.~Paverd, M.~Russinovich, A.~Salem,
  S.~Tople, L.~Wutschitz, and S.~Zanella-B{\'e}guelin, ``Securing {AI} agents
  with information-flow control,'' arXiv:2505.23643, 2025.

\bibitem{agent-permissions}
Y.~Wu, K.~Yang, F.~Roesner, T.~Kohno, N.~Zhang, and U.~Iqbal, ``Towards
  automating data access permissions in {AI} agents,'' in \emph{Proceedings of
  the IEEE Symposium on Security and Privacy (S\&P)}, 2026.

\bibitem{dia}
\BIBentryALTinterwordspacing
{The Browser Company}, ``{Dia}: The browser for better days at work,''
  diabrowser.com, 2026. [Online]. Available: \url{https://www.diabrowser.com/}
\BIBentrySTDinterwordspacing

\bibitem{playwright}
\BIBentryALTinterwordspacing
{Microsoft}, ``{Playwright},'' Playwright Documentation, 2026. [Online].
  Available: \url{https://playwright.dev/}
\BIBentrySTDinterwordspacing

\bibitem{cdp}
\BIBentryALTinterwordspacing
{Google}, ``{Chrome DevTools Protocol},'' Chrome DevTools Documentation, 2026.
  [Online]. Available:
  \url{https://chromedevtools.github.io/devtools-protocol/}
\BIBentrySTDinterwordspacing

\bibitem{chrome-agentic}
\BIBentryALTinterwordspacing
N.~Parker, ``Architecting security for agentic capabilities in {Chrome},''
  Google Security Blog, Dec. 2025. [Online]. Available:
  \url{https://blog.google/security/architecting-security-for-agentic/}
\BIBentrySTDinterwordspacing

\bibitem{edge-copilot}
\BIBentryALTinterwordspacing
S.~Lyndersay, ``Meet {Copilot} mode in {Edge}: Your {AI} browser,'' Microsoft
  Edge Blog, Oct. 2025. [Online]. Available:
  \url{https://blogs.windows.com/msedgedev/2025/10/23/meet-copilot-mode-in-edge-your-ai-browser/}
\BIBentrySTDinterwordspacing

\bibitem{whatwg-dom}
\BIBentryALTinterwordspacing
{WHATWG}, ``{DOM} living standard,'' Web Hypertext Application Technology
  Working Group, 2026. [Online]. Available: \url{https://dom.spec.whatwg.org/}
\BIBentrySTDinterwordspacing

\bibitem{wai-aria}
\BIBentryALTinterwordspacing
J.~Diggs, J.~Nurthen, M.~Cooper, and C.~MacLeod, ``Accessible rich internet
  applications ({WAI-ARIA}) 1.2,'' W3C Recommendation, Jun. 2023. [Online].
  Available: \url{https://www.w3.org/TR/wai-aria-1.2/}
\BIBentrySTDinterwordspacing

\bibitem{webarena}
S.~Zhou, F.~F. Xu, H.~Zhu, X.~Zhou, R.~Lo, A.~Sridhar, X.~Cheng, T.~Ou,
  Y.~Bisk, D.~Fried, U.~Alon, and G.~Neubig, ``{WebArena}: A realistic web
  environment for building autonomous agents,'' in \emph{Proceedings of the
  International Conference on Learning Representations (ICLR)}, 2024.

\bibitem{roth-csp}
S.~Roth, T.~Barron, S.~Calzavara, N.~Nikiforakis, and B.~Stock, ``Complex
  security policy? {A} longitudinal analysis of deployed content security
  policies,'' in \emph{Proceedings of the Network and Distributed System
  Security Symposium (NDSS)}, 2020.

\bibitem{stock-domxss}
B.~Stock, S.~Lekies, T.~Mueller, P.~Spiegel, and M.~Johns, ``Precise
  client-side protection against {DOM}-based cross-site scripting,'' in
  \emph{Proceedings of the USENIX Security Symposium}, 2014.

\bibitem{ucm}
K.~Nikoli{\'c}, E.~Zverev, J.~Rando, M.~Jagielski, E.~Debenedetti, and
  F.~Tram{\`e}r, ``Untrusted content masking for web agents with security
  guarantees,'' arXiv:2607.05277, Jul. 2026.

\bibitem{prismata}
C.~Villa, A.~E. Ozdarendeli, S.~Tan, and R.~A. Popa, ``{Prismata}: Confining
  cross-site prompt injection in web agents,'' arXiv:2607.08147, Jul. 2026.

\bibitem{pagegraph}
\BIBentryALTinterwordspacing
{Brave}, ``{PageGraph},'' Brave Browser Wiki, 2026. [Online]. Available:
  \url{https://github.com/brave/brave-browser/wiki/PageGraph}
\BIBentrySTDinterwordspacing

\bibitem{iqbalAIES24}
U.~Iqbal, T.~Kohno, and F.~Roesner, ``{LLM} platform security: Applying a
  systematic evaluation framework to {OpenAI}'s {ChatGPT} plugins,'' in
  \emph{Proceedings of the AAAI/ACM Conference on AI, Ethics, and Society
  (AIES)}, 2024.

\bibitem{psl}
\BIBentryALTinterwordspacing
{Mozilla}, ``{Public Suffix List},'' publicsuffix.org, 2026. [Online].
  Available: \url{https://publicsuffix.org/}
\BIBentrySTDinterwordspacing

\bibitem{tranco}
V.~L. Pochat, T.~V. Goethem, S.~Tajalizadehkhoob, M.~Korczy{\'n}ski, and
  W.~Joosen, ``Tranco: A research-oriented top sites ranking hardened against
  manipulation,'' in \emph{Proceedings of the Network and Distributed System
  Security Symposium (NDSS)}, 2019.

\bibitem{curlie}
\BIBentryALTinterwordspacing
{Curlie Project}, ``{Curlie},'' curlie.org, 2026. [Online]. Available:
  \url{https://curlie.org/}
\BIBentrySTDinterwordspacing

\bibitem{osworldhuman}
R.~Abhyankar, Q.~Qi, and Y.~Zhang, ``{OSWorld-Human}: Benchmarking the
  efficiency of computer-use agents,'' in \emph{Proceedings of the Conference
  on Machine Learning and Systems (MLSys)}, 2026.

\bibitem{visiblev8}
J.~Jueckstock and A.~Kapravelos, ``{VisibleV8}: In-browser monitoring of
  {JavaScript} in the wild,'' in \emph{Proceedings of the Internet Measurement
  Conference (IMC)}, 2019.

\bibitem{google-migrations}
S.~Nikolov, D.~Codecasa, A.~Sjovall, M.~Tabachnyk, S.~Chandra, S.~Taneja, and
  C.~Ziftci, ``How is {Google} using {AI} for internal code migrations?'' in
  \emph{Proceedings of the IEEE/ACM International Conference on Software
  Engineering: Software Engineering in Practice (ICSE-SEIP)}, 2025.

\bibitem{cellmate}
L.~Meng, H.~Feng, I.~Shumailov, and E.~Fernandes, ``{ceLLMate}: Sandboxing
  browser {AI} agents,'' arXiv:2512.12594, Dec. 2025.

\end{thebibliography}
\end{document}